\documentclass[showpacs,preprintnumbers,amsmath,amssymb,prb]{revtex4}

\usepackage{graphicx}
\usepackage{dcolumn}
\usepackage{bm}

\begin{document}

\title{Bond-operator analytical approach for the $t$--$J$ model}

\author{A.\ V.\ Syromyatnikov}
\email{asyromyatnikov@yandex.ru}
\affiliation{Petersburg Nuclear Physics Institute named by B.P.Konstantinov of National Research Center "Kurchatov Institute", Gatchina 188300, Russia}

\date{\today}

\begin{abstract}

We present a bond-operator theory (BOT) for analytical consideration of the $t$--$J$ model and its extensions with longer-rage hopping terms. This technique is based on previously suggested representation of electron operators via localized spins $1/2$ and spinless Fermi-operators of mobile holons which requires no constraint between them and which provides an exact reformulation of the $t$--$J$ model on a convenient ground. We introduce then a representation of operators of spins, holons, and electrons in (extended) unit cell containing several lattice sites via a rich zoo of Bose- and Fermi-operators acting in the Hilbert space of all quantum states of the {\it whole} unit cell. The proposed BOT provides a regular expansion of physical quantities in powers of $1/n$ using conventional diagrammatic technique, where $n\ge1$ is the maximum number of introduced quasiparticles (bosons and fermions) which can occupy a unit cell. The suggested representation reproduces the commutation algebra of all operators at any $n>0$ and allows to consider both magnetically ordered and disordered phases. 
Some elementary excitations described in the BOT by separate bosons or fermions appear in common approaches as bound states of conventional quasiparticles. In particular, there are two-hole bound states (Cooper pairs of two holes) which are described within the BOT by separate bosons.
To demonstrate the capabilities of the BOT, we discuss in detail properties of the system containing no more than two holes (magnetic polarons) in the particular case of the square lattice with two lattice sites in the (magnetic) unit cell. Although the expansion parameter $1/n$ is not small in the physically meaningful case of $n=1$, we demonstrate that one obtains a good {\it quantitative} agreement with previous numerical findings at $n=1$ even in the first order in $1/n$ after taking into account a few simple diagrams. Self-consistent calculations in the first order in $1/n$ bring our results to a very good {\it quantitative} agreement with previous numerical findings of the ground state energy, the staggered magnetization, and spectra of magnons, polarons, and the lowest-energy two-hole bound states. 

\end{abstract}


\maketitle

\section{Introduction}

The one-band $t$--$J$ model on the square lattice has been extensively studied since the early days of high-temperature superconductivity research \cite{dagotorev,leerev} because it captures essential low-energy physics of hole-doped cuprates as proposed in Refs.~\cite{zhang,anderson}. Its Hamiltonian
\begin{eqnarray}
\label{ham0}
	{\cal H} &=& -t \sum_{\langle p,j\rangle \sigma} 
	\left(
	c^\dagger_{p\sigma} c^{}_{j\sigma}
	+ c^\dagger_{j\sigma} c^{}_{p\sigma} \right)
	+
	J\sum_{\langle p,j\rangle} \left({\bf s}_p{\bf s}_j-\frac14n_pn_j\right),\\
\label{cons}
	n_j &=& \sum_\sigma n_{j\sigma} = \sum_\sigma c^\dagger_{j\sigma} c^{}_{j\sigma} 
	\le 1 \qquad \mbox{for $\forall j$},
\end{eqnarray}
where $t,J>0$, $\sigma=\pm\frac12$, ${\bf s}_j = \frac12 \sum_{\alpha\beta}c^\dagger_{j\alpha}\mbox{\boldmath $\tau$}^{}_{\alpha\beta}c^{}_{j\beta}$, and $\mbox{\boldmath $\tau$}$ is the Pauli vector, incorporates the strongly correlated nature of cuprates through constraint \eqref{cons} excluding simultaneous occupation of a site by two electrons. This constraint brings the system to the Mott insulating phase at half-filling in which case there is one electron per site and Hamiltonian \eqref{ham0} is reduced to the spin-$\frac12$ Heisenberg antiferromagnet whose properties are generally well understood. \cite{monous} Away from the half-filling, carriers arise in the system showing an unconventional superconducting behavior with an important but not completely clear role of short-range antiferromagnetic correlations behind it. \cite{Keimer2015,dagotorev,ogatarev,scalrev,iztj}

Although the $t$--$J$ model and its extensions with longer-range hopping terms appeared to be successful in explaining some characteristic low-temperature properties of cuprates including some features of their rich phase diagram, a comprehensive understanding of its properties is still being sought. \cite{leerev,ogatarev,grosrev} The analytical treatment of the problem is greatly complicated by the absence of a small parameter in the most interesting regime of $J/t=0.3-0.4$ which is relevant to cuprates. Besides, the constraint of the $t$--$J$ model is taken into account only approximately within many approaches that implies an unclear accuracy of theoretical discussions. Then, no controlled solution is known and most successful analytical treatments are based on mean-field considerations withing various slave-boson and slave-fermion formalisms. \cite{leerev,ogatarev,grosrev,iztj} In this situation, numerical methods acquire a particularly important role in the development of which great progress has been made in last decades. In particular, numerical results reproduce the experimentally observed $d$-symmetry of the superconducting order parameter and show that quite many phases have close energies at small temperature. \cite{ogatarev,grosrev,Keimer2015} However numerical studies cannot solve the problem completely because a good analytical theory is required for the interpretation of many of their findings.

It is interesting to note that the problem of the absence of a small parameter in the $t$--$J$ model on different lattices is largely overcome at half-filling when the system is equivalent to the spin-$\frac12$ Heisenberg antiferromagnet with the magnetically ordered ground state. Probably the most popular, simple, and useful approach for ordered nonconducting magnets is the spin-wave theory which is based on the bosonic Holstein-Primakoff spin transformation \cite{hp} and which allows to represent expressions for observable quantities as series in powers of $1/s$, where $s$ is the spin value. Although $s\sim1$ in practice and the expansion parameter $1/s$ is not small, these series converge surprisingly fast in many cases far from critical points so that first $1/s$ terms frequently provide the main quantitative renormalization of observables even in two-dimensional systems with $s=1/2$. \cite{monous} The origin of such a remarkable property is essentially unclear (taking into account also that momenta of summation $k\sim1$ give the main contribution in any order in $1/s$). Apparently, the great role plays here by the good choice of the starting point (the classical limit) and the reproduction of the spin commutation algebra by the Holstein-Primakoff transformation. A drawback of the spin-wave theory is its inability to treat quantitatively "complex" excitations known as bound states of some number of magnons: it is normally impossible to find $1/s$ series for their spectra because one has to take into account unknown infinite series of diagrams. Besides, it has been demonstrated recently both numerically and experimentally that due to strong quantum fluctuations the spin-wave theory cannot describe even qualitatively some aspects of short-wavelength spin dynamics in spin-$\frac12$ antiferromagnets on square and triangular lattices (see discussion in Refs.~\cite{ibot,itri,itrih,itrij1j2,iboth,aktersky} and references therein). 

These surprising numerical and experimental findings in spin-$\frac12$ antiferromagnets have been described in Refs.~\cite{ibot,itri,itrih,itrij1j2,iboth,aktersky} using the bond-operator theory (BOT) suggested in Ref.~\cite{ibot}. This analytical technique is close in spirit to the spin-wave theory but the bosonic spin representation is built in it for all spins in the (extended) unit cell (containing several number of spins) using quantum states of the whole unit cell. The suggested representation contains a rich zoo of bosons some of which describe well known magnons and other bosons describe "complex" excitations which could appear in conventional approaches as bound states of several magnons. There is a formal parameter $n$, the maximum number of quasiparticles which can occupy a unit cell, whose role in this representation is analogous to the spin value $s$ in the spin-wave theory: all observable quantities can be expressed as series in powers of $1/n$ using the conventional diagrammatic technique. Although physical results correspond only to $n=1$, we have found that in many cases first $1/n$ corrections bring our findings at $n=1$ to a quantitative agreement with numerical and experimental results (just as the spin-wave theory does in many cases at $s\sim1$). Besides, the BOT is able to describe magnetically ordered and disordered phases (as well as transitions between them) and provides a simple way to find spectra of some "complex" spin excitations as $1/n$-series by calculating a few simple diagrams (these bound states of several magnons are described in the BOT by separate bosons). In particular, we demonstrate in Ref.~\cite{icup} that anomalies in the Raman spectra in the $B_{1g}$ geometry, in the resonant inelastic x-ray scattering cross section, and in the infrared optical absorption spectra in parent compounds of high-$T_c$ cuprates are produced by the Higgs mode and by another spin-0 quasiparticle named "singlon" which are described in the BOT by separate bosons. Apparently, the origin of the success of the BOT in describing the short-wavelength spin dynamics is that it is built on precise quantum states of the whole unit cell and thus takes into account quite accurately the short-range spin correlations. In addition, the BOT reproduces the commutation algebra of all spin operators in the unit cell at any $n>0$ and it does not mix states from the "physical" and "unphysical" subspaces containing no more than $n$ bosons and more than $n$ bosons in the unit cell, respectively.

\begin{figure}
\includegraphics[scale=0.4]{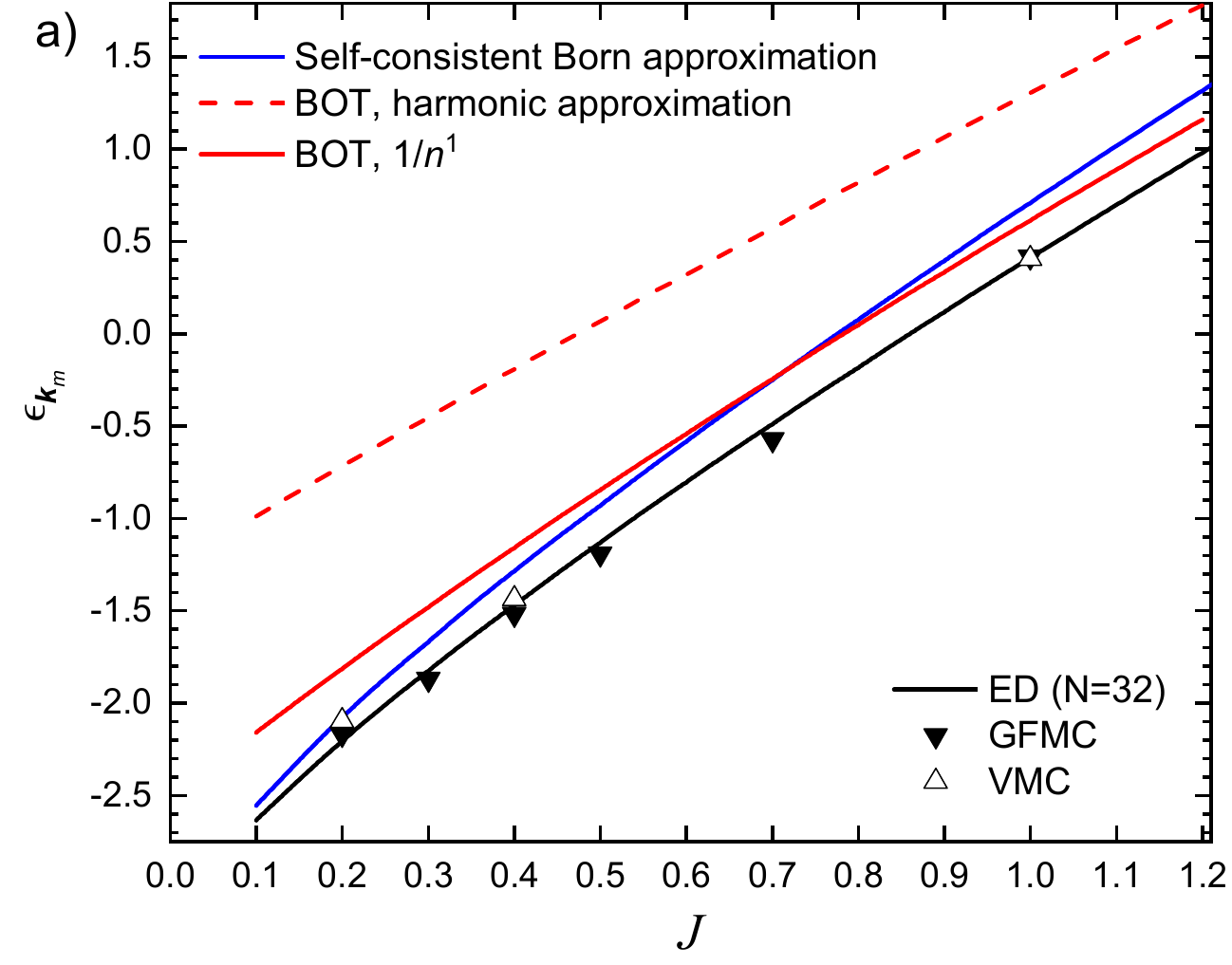}
\includegraphics[scale=0.4]{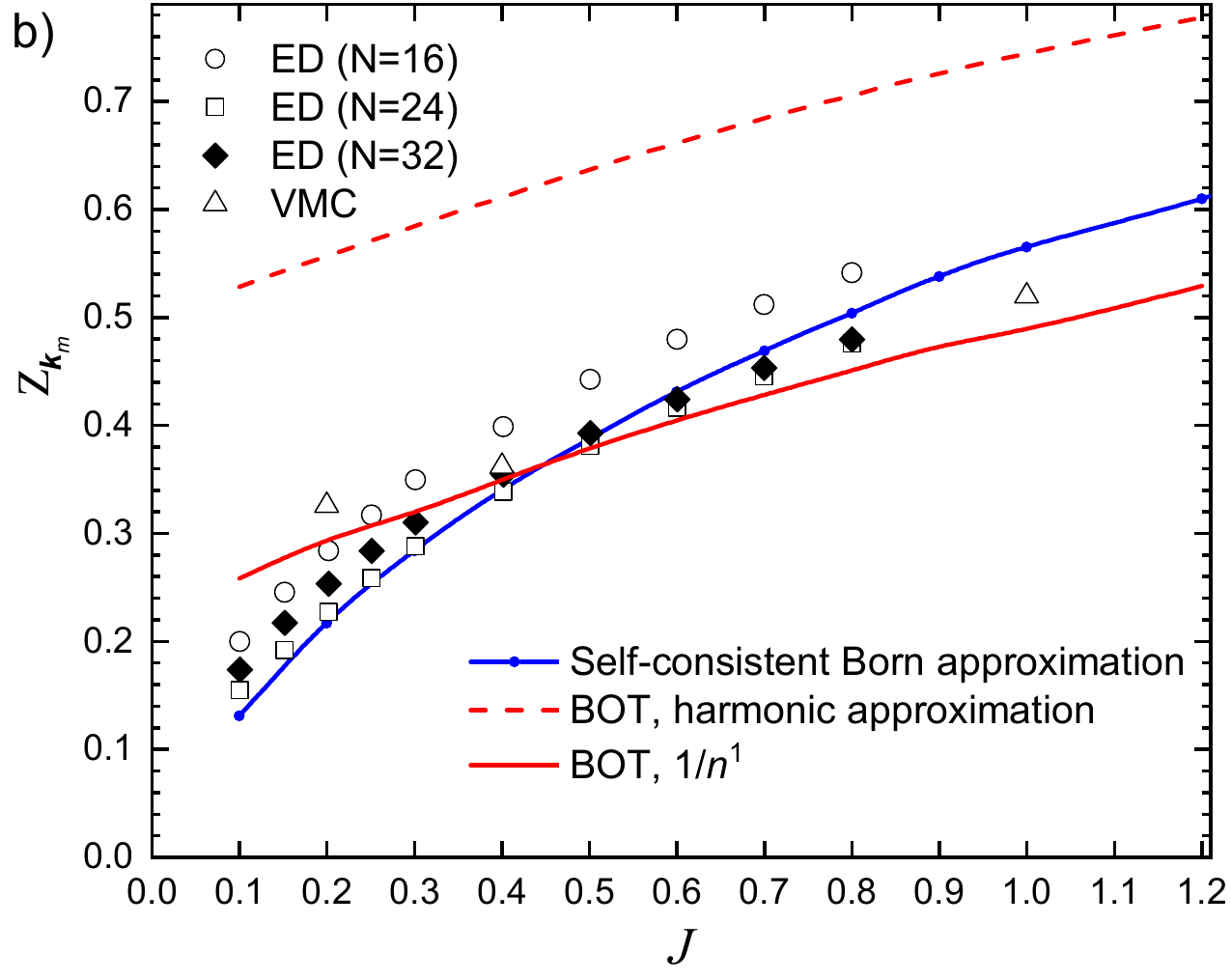}
\includegraphics[scale=0.4]{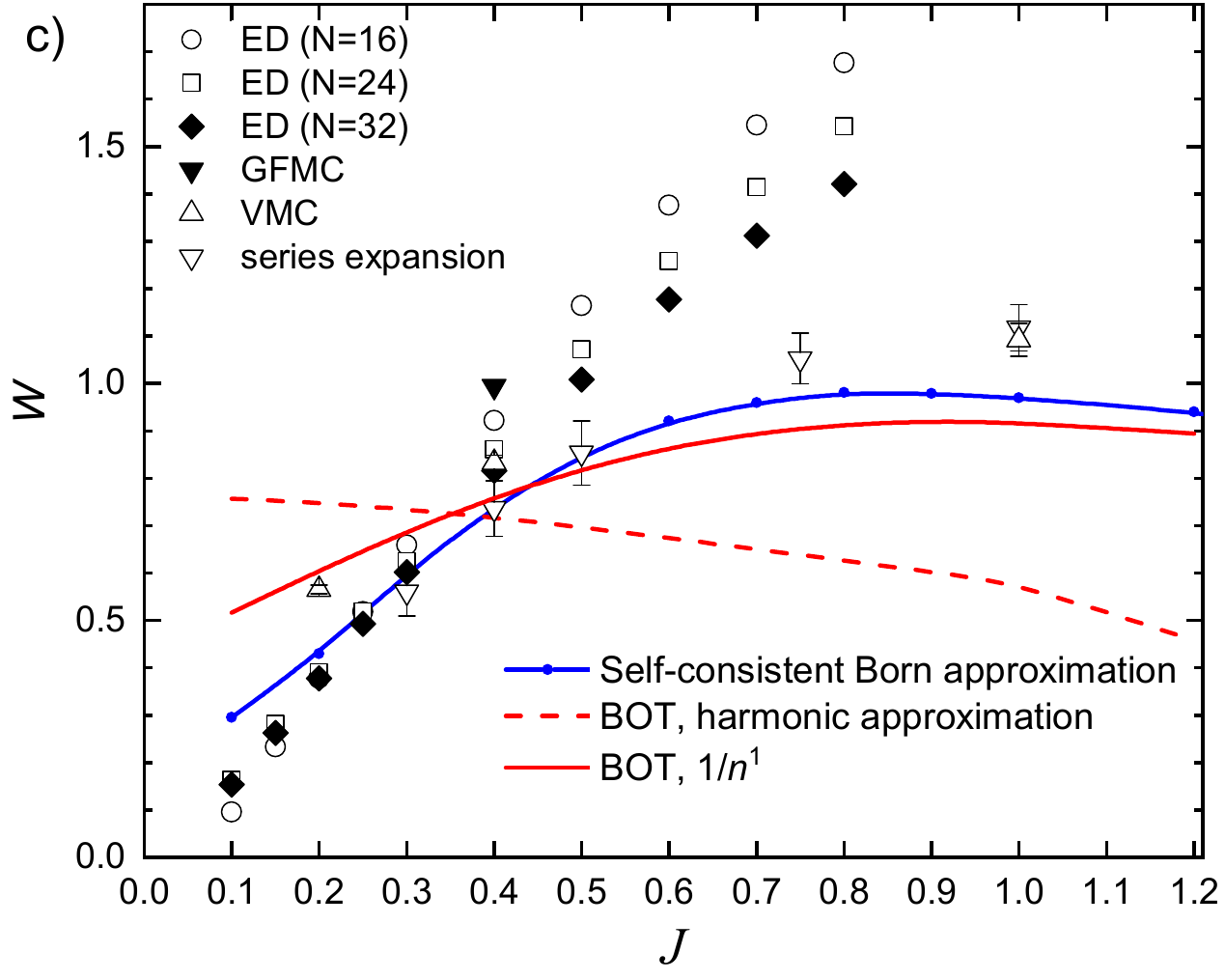}
\caption{
a) Ground state energy of the $t$--$J$ model on the square lattice with one hole counted from the energy of the system without holes (which is given by the spectrum of the magnetic polaron $\epsilon_{\bf k}$ in its minimum at ${\bf k}={\bf k}_m=(\pi/2,\pi/2)$) found using exact diagonalization (ED) of a cluster containing $N=32$ sites (the data are fitted by the formula $-3.24+2.65J^{0.72}$) \cite{ed1}, the self-consistent Born approximation (SCBA) on $16\times16$ cluster (the data are fitted accurately by $J-3.23+2.94J^{0.702}$), \cite{scba1} the Green's function Monte Carlo (GFMC) technique, \cite{gf1,gf2,gf3} variational Monte Carlo (VMC) method, \cite{vmc} and in the BOT (present study). 
b) Spectral weight $Z_{{\bf k=k}_m}$ of the polaron in the electron Green's function. c) Polaron bandwidth $W$. 
We set $t=1$.
\label{ez}}
\end{figure}

The aim of the present paper is to generalize our BOT suggested before for dielectric magnets to the $t$--$J$ model on different lattices away from the half-filling (for hole doping, for definiteness). The hope is that (i) as it was in magnets, $1/n$ series for observables will converge fast in the $t$--$J$ model providing a quantitative description of its dynamics, (ii) the BOT will be just as successful in describing short-range correlations whose important role have been pointed out before in strongly correlated electron systems, and  (iii) the BOT will allow to treat all important low-energy "complex" excitations (some of which might be unknown) and point relevant correlators in which these quasiparticles produce distinct anomalies. To immediately illustrate that at least some of these hopes can come true, we present in Fig.~\ref{ez} the smallest energy of the magnetic polaron $\epsilon_{{\bf k}_m}$ (i.e., the difference of the ground state energy of the system with one hole and without holes), its spectral weight $Z_{{\bf k}_m}$ in the electron Green's function, and the polaron band width $W$ as functions of $J$ which are obtained below on the square lattice. It is seen that these quantities found in the first order in $1/n$ are in a good quantitative agreement with previous findings obtained numerically and within the self-consistent Born approximation (SCBA).

The plan of our consideration is as follows. In the first step, we get rid of the constraint \eqref{cons} of the $t$--$J$ model using the representation of electron operators in Eq.~\eqref{ham0} via localized spins $1/2$ and spinless Fermi operators of holes (holons) which was developed in Refs.~\cite{sb1,sb2}. This approach provides an exact and convenient reformulation of the $t$--$J$ model free of constraints which is described in Sec.~\ref{constr} for completeness. After this procedure, the system takes the form of the Heisenberg antiferromagnet with some number of fermionic holons which interact strongly with spins and can hop from site to site. Then, we illustrate in Sec.~\ref{method2} the main steps of building the bond operator approach with the simplest example of the unit cell containing two lattice sites (i.e., the magnetic unit cell). We present representations of all operators in the model via appropriate Fermi and Bose operators (some of derived representations are given in Appendix~\ref{oprep}). We apply in Sec.~\ref{hafbi} this formalism to consideration of static and dynamical properties of the $t$--$J$ model on the square lattice. We discuss the results in Sec.~\ref{disc} and provide our conclusion and the outlook. In Appendix~\ref{method1}, we use the suggested procedure of constructing representations of operators for one site in the unit cell. We do not consider results of the one-site formalism in detail because our operators representations turned out to be very similar to that derived in Ref.~\cite{chang} from other considerations and because self-consistent calculations are required in the one-site formalism to achieve sufficiently good quantitative agreement with previous numerical results for polaron properties (see Appendix~\ref{method1}).

\section{Representation of electron operators free from constraints}
\label{constr}

For the sake of completeness, we discuss in this section the representation of electron operators in the $t$--$J$ model \eqref{ham0} via localized spins $1/2$ and spinless Fermi-operators of holes (holons) which was originally proposed in Ref.~\cite{sb1} and refined in Ref.~\cite{sb2} and which allows to get rid of constraint \eqref{cons}. Similar trick was used also in Ref.~\cite{giniyat}.

Due to constraint \eqref{cons}, the Hilbert space at any site $j$ of the $t$--$J$ model can be considered as limited by states $|j\sigma\rangle$ and $|j0\rangle$ corresponding, respectively, to the site occupied by one electron with projection $\sigma$ and to the empty site. Let us introduce a tensor product space ${\cal H}_j^s\otimes{\cal H}_j^h$ at the lattice site $j$, where ${\cal H}_j^h$ consists of states $|j1\rangle_h$ and $|j0\rangle_h$ with and without a hole, respectively, and ${\cal H}_j^s$ is a spin-$\frac12$ space built on states 
$|j\!\uparrow\rangle_s$ and $|j\!\downarrow\rangle_s$ ($\sigma=\frac12$ and $\sigma=-\frac12$ are denoted by $\uparrow$ and $\downarrow$, respectively). Let us bring states $|j\sigma\rangle$ and $|j0\rangle$ into correspondence with three states from ${\cal H}_j^s\otimes{\cal H}_j^h$ as follows:
\begin{equation}
\label{mapst}
	|j\sigma\rangle \mapsto |j\sigma\rangle_s |j0\rangle_h,
	\qquad
	|j0\rangle \mapsto |jS\rangle_s |j1\rangle_h,
\end{equation}
where 
\begin{eqnarray}
\label{sst}
	|jS\rangle_s &=& u_1|j\!\uparrow\rangle_s + u_2|j\!\downarrow\rangle_s,\\
\label{u1}
	u_1 &=& \cos\theta, \\
\label{u2}
	u_2 &=& e^{i\phi}\sin\theta
\end{eqnarray}
and $\phi$ and $\theta$ are some real parameters. The remaining fourth state
\begin{equation}
\label{stst}
	\left|j\overline{S}\right\rangle_s|j1\rangle_h = 
	\Bigl( u_2^*|j\!\uparrow\rangle_s - u_1^*|j\!\downarrow\rangle_s \Bigr) |j1\rangle_h 
\end{equation}
from ${\cal H}_j^s\otimes{\cal H}_j^h$ and values of $\phi$ and $\theta$ are discussed below. Introducing the spin-$\frac12$ operator ${\bf S}_j$ acting in ${\cal H}_j^s$ and the fermionic holon operator $h_j$ acting in ${\cal H}_j^h$ as
\begin{equation}
\label{hdef}
	h_j^\dagger |j0\rangle_h = |j1\rangle_h,
	\qquad
	h_j |j1\rangle_h = |j0\rangle_h
\end{equation}
one can bring these operators into correspondence with the electron operators in Hamiltonian \eqref{ham0} using Eqs.~\eqref{mapst} as follows:
\begin{eqnarray}
\label{cop}
	c_{j\uparrow} &\mapsto& {\mathfrak c}_{j\uparrow} = h_j^\dagger \left( u_1S^+_jS^-_j + u_2S^-_j \right),\nonumber\\
	c_{j\uparrow}^\dagger &\mapsto& {\mathfrak c}_{j\uparrow}^\dagger = h_j \left( u_1^*S^+_jS^-_j + u_2^*S^+_j \right),\\
	c_{j\downarrow} &\mapsto& {\mathfrak c}_{j\downarrow} = h_j^\dagger \left( u_1S^+_j + u_2S^-_jS^+_j \right),\nonumber\\
	c_{j\downarrow}^\dagger &\mapsto& {\mathfrak c}_{j\downarrow}^\dagger = h_j \left( u_1^*S^-_j + u_2^*S^-_jS^+_j \right).\nonumber
\end{eqnarray}
Indeed, according to Eqs.~\eqref{mapst}, \eqref{sst}, and \eqref{hdef}, operators $c_{j\sigma}$ and $c_{j\sigma}^\dagger$ act on states $|j\sigma\rangle$ and $|j0\rangle$ exactly as operators ${\mathfrak c}_{j\sigma}$ and ${\mathfrak c}_{j\sigma}^\dagger$ act on the corresponding states from ${\cal H}_j^s\otimes{\cal H}_j^h$. The advantage of this reformulation is that no constraint like Eq.~\eqref{cons} is needed which restricts the Hilbert space. This is because the Hamiltonian and observable quantities are expressed in terms of operators $c_{j\sigma}$ and $c_{j\sigma}^\dagger$ (${\mathfrak c}_{j\sigma}$ and ${\mathfrak c}_{j\sigma}^\dagger$) and the action of ${\mathfrak c}_{j\sigma}$ and ${\mathfrak c}_{j\sigma}^\dagger$ on state \eqref{stst} not involving in Eq.~\eqref{mapst} gives zero as it can be checked straightforwardly using Eqs.~\eqref{cop}: 
\begin{equation}
	{\mathfrak c}_{j\sigma}\left|j\overline{S}\right\rangle_s|j1\rangle_h
	= {\mathfrak c}^\dagger_{j\sigma}\left|j\overline{S}\right\rangle_s|j1\rangle_h
	=0.
\end{equation}
The disappearance of constraint \eqref{cons} also occurs at the formal level:
\begin{equation}
1\ge \sum_\sigma c^\dagger_{j\sigma} c^{}_{j\sigma} \quad \mapsto \quad
1\ge \sum_\sigma {\mathfrak c}^\dagger_{j\sigma} {\mathfrak c}^{}_{j\sigma} = h_j h_j^\dagger = 1-h_j^\dagger h_j,
\end{equation}
where the latter inequality which reads as $h_j^\dagger h_j\ge0$ is satisfied automatically. Then, there is no constraint after representation \eqref{cop}.

According to Eqs.~\eqref{cop}, initial spin operators ${\bf s}_j = \frac12 \sum_{\alpha\beta}c^\dagger_{j\alpha}\mbox{\boldmath $\tau$}^{}_{\alpha\beta}c^{}_{j\beta}$ transform as
\begin{equation}
\label{s}
	{\bf s}_j \mapsto h_j h_j^\dagger {\bf S}_j
\end{equation}
and Hamiltonian \eqref{ham0} acquires the form of the Heisenberg antiferromagnet with fermionic holons which interact with the localized spins and can hop from site to site:
\begin{equation}
\label{ham1}
	\widetilde{\cal H} 
	= 
	t \sum_{\langle p,j\rangle} 
	\left( 
	h^\dagger_ph_j \left[ \frac14 + {\bf S}_p{\bf S}_j 
	+ {\bf f}\cdot \left( \frac12{\bf S}_p + \frac12{\bf S}_j - i\left[{\bf S}_p\times{\bf S}_j\right] \right)
	\right] + {\rm H.c.}
	\right)
	+
	J\sum_{\langle p,j\rangle} \left(h_ph^\dagger_p{\bf S}_p{\bf S}_jh_jh^\dagger_j
	-\frac14h_ph^\dagger_ph_jh^\dagger_j\right),
\end{equation}
where ${\bf f} = (\sin2\theta\cos\phi,\sin2\theta\sin\phi,\cos2\theta)$ (see Eqs.~\eqref{u1} and \eqref{u2}). Notice that Eq.~\eqref{ham1} is reduced to the Heisenberg antiferromagnet when there are no holes in the system (i.e., at half filling of the initial model \eqref{ham0}). Assuming that spins ${\bf S}_j$ has the same time-reversal symmetry as the initial spins ${\bf s}_j$, one should impose the condition $|u_1|=|u_2|$ so that $\widetilde{\cal H}$ obeys the time-reversal symmetry as the initial $t$--$J$ model. \cite{sb2} As a result, it was proposed in Ref.~\cite{sb2} to put $\theta=\pi/4$ and to leave $\phi$ as arbitrary in Eqs.~\eqref{u1} and \eqref{u2}. Notice that effective Hamiltonian \eqref{ham1} is invariant under global spin rotations around $\bf f$. Then, there should be a Goldstone magnetic excitation if the rotational symmetry is spontaneously broken in the plane perpendicular to $\bf f$.

The range of stability of the antiferromagnetic ordering in model \eqref{ham1} against the hole doping $x$ was studied in Ref.~\cite{sb3} in the mean-field approximation. It was shown that an effective magnetic field directed along $\bf f$ arises in the spin subsystem because $h^\dagger_ph_j$ in Eq.~\eqref{ham1} is replaced by its mean value proportional to $x$ during the mean-field decoupling of the spin-holon interaction term (notice also that $\langle h^\dagger_ph_j\rangle = \langle h^\dagger_jh_p\rangle$ and the term with the vector product disappears in Eq.~\eqref{ham1} in the mean-field approximation). Then, the magnetically disordered phase is the forced ferromagnetic state in this theory which arises at a critical value $x_c$. Although values of $x_c$ were found to be in a good quantitative agreement with previous numerical findings for some particular $J$ values, the nature of the disordered state (the forced ferromagnet) was inconsistent with them. \cite{sb3} Apparently, this is why the formalism presented in this section has not been further developed. 

We find it convenient below not to consider Eq.~\eqref{ham1} withing the BOT. Instead, we built in the next section appropriate representations directly for operators \eqref{cop} and \eqref{s} and substitute them to initial Hamiltonian \eqref{ham0}. We obtain a good agreement of our findings with previous numerical results. In particular, we will show in subsequent papers that there is no ferromagnetic mean spin component in the BOT at finite $x$. Besides, we check below by straightforward calculations in the first order in $1/n$ that spectra of all quasiparticles do not depend on $\theta$ and $\phi$ in Eqs.~\eqref{u1} and \eqref{u2}. Then, we set
\begin{equation}
\label{u12}
	u_1=u_2=\frac{1}{\sqrt2}
\end{equation}
below for definiteness.

\section{Bond-operator technique for two sites in the unit cell}
\label{method2}

\begin{figure}
\includegraphics[scale=0.5]{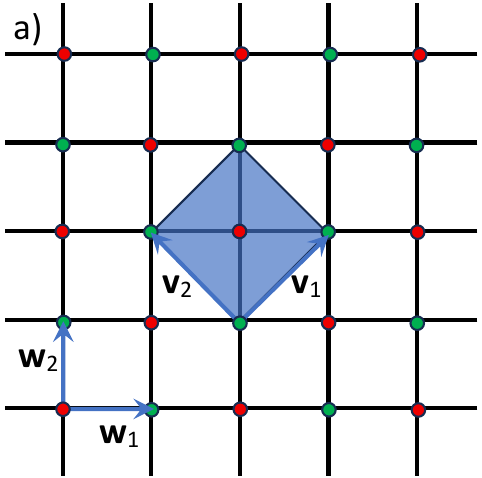}
\includegraphics[scale=0.33]{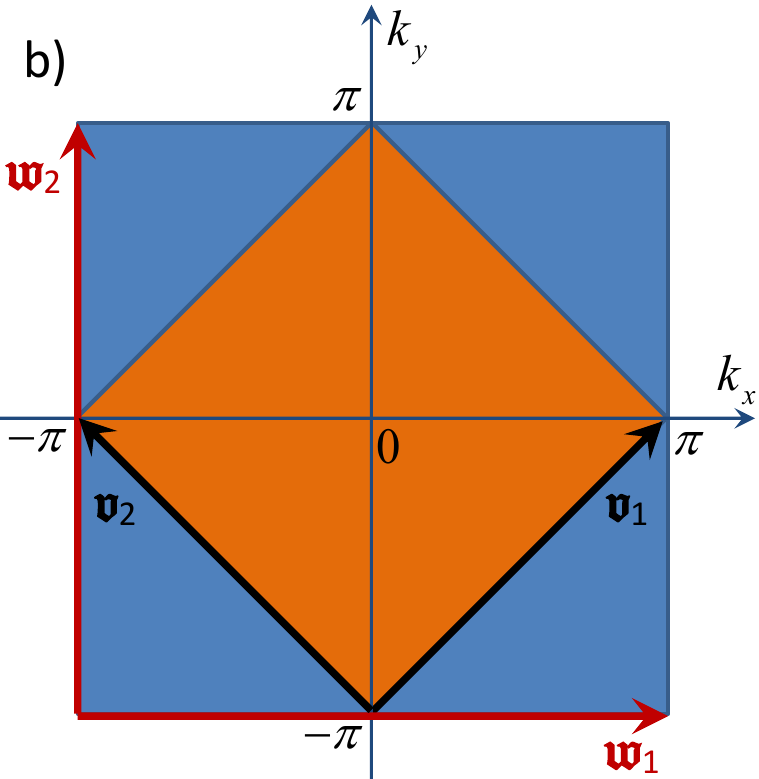}
\caption{(a) Square lattice with two sites in the unit cell highlighted in color. Also shown are corresponding translation vectors ${\bf v}_{1,2}$ as well as translation vectors ${\bf w}_{1,2}$ of the square lattice. (b) Brillouin zones corresponding to ${\bf v}_{1,2}$ (smaller square with respective translation vectors $\mbox{\boldmath $\mathfrak v$}_{1,2}$) and ${\bf w}_{1,2}$ (larger square with respective translation vectors $\mbox{\boldmath $\mathfrak w$}_{1,2}$). The distance between nearest lattice sites is set to be equal to unity.
\label{system}}
\end{figure}

Bearing in mind for definiteness the $t$--$J$ model on the simple square lattice shown in Fig.~\ref{system}(a), we develop in this section the BOT for the unit cell containing a bond with two nearest sites. In the magnetically order state of the $t$--$J$ model on the simple square lattice, the unit cell contains two sites from different magnetic sublattices (see Fig.~\ref{system}(a)). At half-filling, model \eqref{ham0} has been well studied before by various methods (see, e.g., Refs.~\cite{monous} and references therein). We tested in Ref.~\cite{ibot} the two-site variant of the bond-operator formalism on the square-lattice bilayer at half-filling and found a good quantitative agreement with previous numerical and analytical results. We generalize now the technique suggested in Ref.~\cite{ibot} for consideration of the $t$--$J$ model and its extensions away from the half-filling.

States of the $j$-th unit cell can be represented as linear combinations of 16 states $|jS_1S_2\rangle_s|jh_1h_2\rangle_h$, where $S_l=\uparrow,\downarrow$ signify the spin state at $l$-th site of the unit cell and $h_l=0,1$ is the hole occupation number ($l=1,2$). For building representations of operators appearing in the considered model, it is convenient to introduce the vacuum state in the form
\begin{equation}
\label{vac}
|j0\rangle = \Bigl(\cos\alpha\left|j\!\uparrow\downarrow\rangle_s\right.
- \sin\alpha\left|j\!\downarrow\uparrow\rangle_s\right.\Bigr)|j00\rangle_h,
\end{equation}
where $\alpha$ is a real parameter. It is seen that the vacuum $|j0\rangle$ is a singlet state at $\alpha=\pi/4$ whereas the N\'eel order
(i.e., $\langle j0|S_{1j}^z|j0\rangle = -\langle j0|S_{2j}^z|j0\rangle \ne0$)
arises when $\sin\alpha\ne\cos\alpha$. Parameter $\alpha$ allows to connect smoothly the N\'eel ordered phase and the singlet state which arises when one weaken sufficiently all couplings between spins from different unit cells. Then, $\alpha$ depends on the inter-bond interaction and it is found below for the isotropic square lattice. 

\begin{table}
\caption{Definition of 7 Bose operators $a_{qj}$, $b_{pj}$ and 8 Fermi operators $d_{pj}$, $e_{pj}$ ($q=1,2,3$ and $p=1,2,3,4$) in the $j$-th unit cell which act on the vacuum state $|j0\rangle$ given by Eq.~\eqref{vac}. To lighten notation, index $j$ is omitted.
\label{statestab}
}
\begin{ruledtabular}
\begin{tabular}{ll}
\multicolumn{2}{c}{Bose operators} \\
\hline
$a_{1}^\dagger |0\rangle = |a_1\rangle = \left|\uparrow\uparrow\rangle_s\right.|00\rangle_h$
 &
$b_{1}^\dagger |0\rangle = |b_1\rangle = \Bigl(\cos\alpha\left|\uparrow\downarrow\rangle_s\right.
- \sin\alpha\left|\downarrow\uparrow\rangle_s\right.\Bigr)|11\rangle_h$
\\
$a_{2}^\dagger |0\rangle = |a_2\rangle = \left|\downarrow\downarrow\rangle_s\right.|00\rangle_h$
 & 
$b_{2}^\dagger |0\rangle = |b_2\rangle = \left|\uparrow\uparrow\rangle_s\right.|11\rangle_h$
\\
$a_{3}^\dagger |0\rangle = |a_3\rangle = \Bigl(\sin\alpha\left|\uparrow\downarrow\rangle_s\right.
+ \cos\alpha\left|\downarrow\uparrow\rangle_s\right.\Bigr)|00\rangle_h$
 & 
$b_{3}^\dagger |0\rangle = |b_3\rangle = \left|\downarrow\downarrow\rangle_s\right.|11\rangle_h$
\\
&
$b_{4}^\dagger |0\rangle = |b_4\rangle = \Bigl(\sin\alpha\left|\uparrow\downarrow\rangle_s\right.
+ \cos\alpha\left|\downarrow\uparrow\rangle_s\right.\Bigr)|11\rangle_h$
 \\
\hline
\multicolumn{2}{c}{Fermi operators} \\
\hline
$d_{1}^\dagger |0\rangle = |d_1\rangle = \Bigl(\cos\alpha\left|\uparrow\downarrow\rangle_s\right.
- \sin\alpha\left|\downarrow\uparrow\rangle_s\right.\Bigr)|10\rangle_h$
&
$e_{1}^\dagger |0\rangle = |e_1\rangle = \Bigl(\cos\alpha\left|\uparrow\downarrow\rangle_s\right.
- \sin\alpha\left|\downarrow\uparrow\rangle_s\right.\Bigr)|01\rangle_h$\\
$d_{2}^\dagger |0\rangle = |d_2\rangle = \left|\uparrow\uparrow\rangle_s\right.|10\rangle_h$
 &
$e_{2}^\dagger |0\rangle = |e_2\rangle = \left|\uparrow\uparrow\rangle_s\right.|01\rangle_h$
 \\
$d_{3}^\dagger |0\rangle = |d_3\rangle = \left|\downarrow\downarrow\rangle_s\right.|10\rangle_h$
 &
$e_{3}^\dagger |0\rangle = |e_3\rangle = \left|\downarrow\downarrow\rangle_s\right.|01\rangle_h$
 \\
$d_{4}^\dagger |0\rangle = |d_4\rangle = \Bigl(\sin\alpha\left|\uparrow\downarrow\rangle_s\right.
+ \cos\alpha\left|\downarrow\uparrow\rangle_s\right.\Bigr)|10\rangle_h$
 &
$e_{4}^\dagger |0\rangle = |e_4\rangle = \Bigl(\sin\alpha\left|\uparrow\downarrow\rangle_s\right.
+ \cos\alpha\left|\downarrow\uparrow\rangle_s\right.\Bigr)|01\rangle_h$
 \\
\end{tabular}
\end{ruledtabular}
\end{table}

Let us introduce now 7 Bose operators $a_{qj}^\dagger$, $b_{pj}^\dagger$ and 8 Fermi operators $d_{pj}^\dagger$, $e_{pj}^\dagger$ ($q=1,2,3$ and $p=1,2,3,4$) which create the remaining 15 states of the unit cell from the vacuum state $|j0\rangle$ as it is defined in Table~\ref{statestab}. Notice that they create states with even and odd number of holons, respectively. Operators $a_{1j}^\dagger$ and $a_{2j}^\dagger$ describe spin-1 excitations which are ordinary magnons in the ordered phase and spin-1 triplons in the disordered (singlet) state whereas $a_{3j}^\dagger$ describes a spin-0 excitation which is the amplitude (Higgs) mode in the ordered phase and the spin-0 triplon in the singlet state (see below and Ref.~\cite{ibot} for extra detail about magnetic excitations at half filling). States $|jd_p\rangle$ and $|je_p\rangle$ describe holons dressed in different magnetic coats. Bosonic operators $b_{pj}^\dagger$ creating two-holon states describe bound states of two holons.

\begin{table}
\caption{Results of action of operators ${\bf S}_{1j}$, ${\bf S}_{2j}$, and $({\bf S}_{1j}{\bf S}_{2j})$ on states $|j0\rangle$ and $|ja_{1,2,3}\rangle$ (see Eq.~\eqref{vac} and Table~\ref{statestab}). Their action on $|jr_{1,2,3,4}\rangle$, where $r=b,d,e$, is obtained from this table by replacements $|j0\rangle\to|jr_1\rangle$ and $|ja_i\rangle\to|jr_{i+1}\rangle$. Index $j$ is omitted for brevity.
\label{table1}
}
\begin{ruledtabular}
\begin{tabular}{|c|c|c|c|c|c|}
  { }      & $S^z_1$ & $S^z_2$ & $S^+_1$ & $S^+_2$ & $({\bf S}_1{\bf S}_2)$ \\
\hline
$|0\rangle$ & $\displaystyle \frac{\cos2\alpha}{2}|0\rangle + \frac{\sin2\alpha}{2}|a_3\rangle$ 
& $\displaystyle -\frac{\cos2\alpha}{2}|0\rangle - \frac{\sin2\alpha}{2}|a_3\rangle$ 
& $-\sin\alpha|a_1\rangle$
& $\cos\alpha|a_1\rangle$
& $\displaystyle - \frac{1 + 2\sin2\alpha}{4} |0\rangle + \frac{\cos 2\alpha}{2}|a_3\rangle$  \\
$|a_1\rangle$ & $\displaystyle \frac12|a_1\rangle$ 
& $\displaystyle \frac12|a_1\rangle$ 
& $0$
& $0$
& $\displaystyle \frac14|a_1\rangle$ { } \\
$|a_2\rangle$ & $\displaystyle -\frac12|a_2\rangle$ 
& $\displaystyle -\frac12|a_2\rangle$  
& $\displaystyle \cos\alpha|0\rangle+\sin\alpha|a_3\rangle$ 
& $\displaystyle -\sin\alpha|0\rangle+\cos\alpha|a_3\rangle$ 
& \qquad $\displaystyle \frac14|a_2\rangle$  \\
$|a_3\rangle$ & $\displaystyle \frac{\sin2\alpha}{2}|0\rangle - \frac{\cos2\alpha}{2}|a_3\rangle $ 
& $\displaystyle - \frac{\sin2\alpha}{2}|0\rangle + \frac{\cos2\alpha}{2}|a_3\rangle $ 
& $\cos\alpha|a_1\rangle$
& $\sin\alpha|a_1\rangle$
& $\displaystyle \frac{\cos 2\alpha}{2}|0\rangle - \frac{1 - 2\sin2\alpha}{4} |a_3\rangle $ 
\end{tabular}
\end{ruledtabular}
\end{table}

Representation of operators ${\bf S}_{lj}$, $({\bf S}_{1j}{\bf S}_{2j})$, $[{\bf S}_{1j}\times{\bf S}_{2j}]$, $h_{lj}$, and ${\mathfrak c}_{lj\sigma}$ (as well as other operators composed from these) via operators $a_{qj}$, $b_{pj}$, $d_{pj}$, and $e_{pj}$ can be built according to their action on states $|ja_q\rangle$, $|jb_p\rangle$, $|jd_p\rangle$, and $|je_p\rangle$ introduced in Table~\ref{statestab}. We demonstrate this procedure now by the example of $h_{lj}$, ${\bf S}_{lj}$, $h_{1j}^\dagger h_{2j}$, and $({\bf S}_{1j}{\bf S}_{2j})$ (the action of spin operators on basis states is summarized in Table~\ref{table1}). We propose the following representation for them:
\begin{subequations}
\label{trans}
\begin{eqnarray}
\label{s1+}
S_{1}^+ &=& \left(S_{1}^-\right)^\dagger = \cos\alpha P a_{2} - \sin\alpha a_{1}^\dagger P 
+ \cos\alpha \left( a_{1}^\dagger a_{3} + \sum_{r=b,d,e} \left( r_{1}^\dagger r_{3} + r_{2}^\dagger r_{4} \right) \right)\nonumber\\
&&{} 
+ \sin\alpha \left( a_{3}^\dagger a_{2} + \sum_{r=b,d,e} \left( r_{4}^\dagger r_{3} - r_{2}^\dagger r_{1} \right) \right),\\
\label{s1z}
S_{1}^z &=& n\frac{\cos2\alpha}{2} 
+\frac{\sin 2\alpha}{2} \left(P a_{3} + a_{3}^\dagger P\right) 
+ \frac12\left( a_{1}^\dagger a_{1} - a_{2}^\dagger a_{2} 
+ \sum_{r=b,d,e} \left( r_{2}^\dagger r_{2} - r_{3}^\dagger r_{3} \right) \right)
\\
&&{}+\frac{\sin 2\alpha}{2} \sum_{r=b,d,e} \left(r_{1}^\dagger r_{4} + r_{4}^\dagger r_{1}\right)
- \frac{\cos2\alpha}{2}\left( a_{1}^\dagger a_{1} + a_{2}^\dagger a_{2} + 2a_{3}^\dagger a_{3} 
+ \sum_{r=b,d,e} \left( r_{2}^\dagger r_{2} + r_{3}^\dagger r_{3} + 2r_{4}^\dagger r_{4} \right) \right),\nonumber\\
S_{2}^+ &=& \left(S_{2}^-\right)^\dagger = \cos\alpha a_{1}^\dagger P - \sin\alpha P a_{2} 
+ \cos\alpha \left( a_{3}^\dagger a_{2} + \sum_{r=b,d,e} \left( r_{2}^\dagger r_{1} + r_{4}^\dagger r_{3} \right) \right)\nonumber\\
&&{} + \sin\alpha \left( a_{1}^\dagger a_{3} + \sum_{r=b,d,e} \left( r_{2}^\dagger r_{4} - r_{1}^\dagger r_{3} \right) \right),\\
\label{s2z}
S_{2}^z &=& -n\frac{\cos2\alpha}{2} 
-\frac{\sin 2\alpha}{2} \left(P a_{3} + a_{3}^\dagger P\right) 
+ \frac12\left( a_{1}^\dagger a_{1} - a_{2}^\dagger a_{2} 
+ \sum_{r=b,d,e} \left( r_{2}^\dagger r_{2} - r_{3}^\dagger r_{3} \right) \right)
\\
&&{}-\frac{\sin 2\alpha}{2} \sum_{r=b,d,e} \left(r_{1}^\dagger r_{4} + r_{4}^\dagger r_{1}\right)
+ \frac{\cos2\alpha}{2}\left( a_{1}^\dagger a_{1} + a_{2}^\dagger a_{2} + 2a_{3}^\dagger a_{3} 
+ \sum_{r=b,d,e} \left( r_{2}^\dagger r_{2} + r_{3}^\dagger r_{3} + 2r_{4}^\dagger r_{4} \right) \right),\nonumber\\
\label{ss}
\left({\bf S}_{1}{\bf S}_{2}\right) &=& - n^2\frac{1+2\sin2\alpha}{4} 
+ n\frac{\cos 2\alpha}{2} \left(P a_{3} + a_{3}^\dagger P\right) 
+ n\frac{1+\sin2\alpha}{2}\left(a_{1}^\dagger a_{1} + a_{2}^\dagger a_{2}\right) 
+ \frac n2\sum_{r=b,d,e} \left( r_{2}^\dagger r_{2} + r_{3}^\dagger r_{3} \right)\nonumber\\
&&{}
+ n\frac{\cos2\alpha}{2}\sum_{r=b,d,e} \left(r_{1}^\dagger r_{4} + r_{4}^\dagger r_{1}\right)
+ n\frac{\sin2\alpha}{2}\left( 2a_{3}^\dagger a_{3} 
+ \sum_{r=b,d,e} \left( r_{2}^\dagger r_{2} + r_{3}^\dagger r_{3} + 2r_{4}^\dagger r_{4} \right) \right),\\
h_1^\dagger &=& \frac{1}{\sqrt n} \left(d_1^\dagger P + \sum_{i=1}^3 d_{i+1}^\dagger a_i + \sum_{i=1}^4 b_{i}^\dagger e_i \right),\\
h_2^\dagger &=& \frac{1}{\sqrt n} \left(e_1^\dagger P + \sum_{i=1}^3 e_{i+1}^\dagger a_i - \sum_{i=1}^4 b_{i}^\dagger d_i \right),\\
h_1^\dagger h_2^{} &=& \sum_{i=1}^4 d_{i}^\dagger e_i,
\end{eqnarray}
\end{subequations}
where we omit index $j$ for brevity, 
\begin{equation}
\label{proj}
P = \sqrt{n - \sum_{i=1}^3 a_i^\dagger a_i - \sum_{i=1}^4 \sum_{r=b,d,e} r_i^\dagger r_i},
\end{equation}
and $n=1$. It is easy to check using Table~\ref{table1} that if $n=1$ operators in left-hand sides of Eqs.~\eqref{trans} act on basis states defined in Table~\ref{statestab} and Eq.~\eqref{vac} as operators in right-hand sides. It can be verified straightforwardly that for any $\alpha$ and $n>0$ representation \eqref{trans} reproduces the commutation algebra of all operators (i.e., $[S^\sigma_{lj},S^\beta_{qj}]=i\epsilon_{\sigma\beta\gamma}\delta_{lq}S^\gamma_{lj}$, $[h_{lj},{\bf S}_{qj}]=0$, $\{h_{lj}^\dagger,h_{qj}\}=\delta_{lq}$, and $\{h_{lj},h_{qj}\}=0$) and $({\bf S}_{1j}{\bf S}_{2j})$ given by Eq.~\eqref{ss} commutes with ${\bf S}_{1j}+{\bf S}_{2j}$. The only reason to multiply linear in operators terms in Eqs.~\eqref{trans} by $P$ given by Eq.~\eqref{proj} is to ensure the commutation algebra fulfillment. In addition, due to $P$, Eqs.~\eqref{trans} have zero matrix elements between states from the Hilbert subspace with no more than $n$ quasiparticles in the unit cell ("physical" subspace) and states with more than $n$ quasiparticles ("unphysical" subspace). Then, parameter $n$ plays in this formalism the role of maximum number of all introduced quasiparticles which can occupy a unit cell. Although $n$ can be considered arbitrary in all calculations within the suggested formalism, only the case of $n=1$ has the physical meaning. Parameter $\alpha$ involved in Eqs.~\eqref{trans} is to be found as series in $1/n$ by minimization of the ground-state energy (see below).

It should be stressed that the reproduction by Eqs.~\eqref{trans} of the commutation algebra of all operators guarantees the existence of Goldstone excitations in any order in $1/n$ in phases with some spontaneously broken continuous symmetry (in particular, in magnetically ordered and/or superconducting phases). 

An algorithm can be easily formulated to construct a representation like \eqref{trans} for any initial operator $\cal Q$ (expressed via electron operators) from the result of its action on basis states of the unit cell: one should (i) write down the simplest expression containing products of no more than two Bose and/or Fermi operators which reproduce the action of $\cal Q$ on basis states (e.g., using a table like Table~\ref{table1}), (ii) each constant term in the resulting expression should be multiplied by $n$, (iii) each linear term in Bose and Fermi operators should be multiplied by $P$ given by Eq.~\eqref{proj}, (iv) the result for Fermi operators should be divided by $\sqrt n$, and (v) if $\cal Q$ is a product of two operators, its representation should by additionally multiplied by $n$. This algorithm (which can be programmed, e.g., in Mathematica Software) can be easily generalized to the case of more than two sites in the unit cell. 

We find it more convenient and straightforward to derive Bose-Fermi analogs of all relevant operators in the unit cell (including $({\bf S}_{1j}{\bf S}_{2j})$ and $c^\dagger_{1j\sigma} c^{}_{2j\sigma}$) using this procedure: it allows to make all terms in the Hamiltonian containing products of the same number of operators to be of the same order in $1/n$. Corresponding operators representation is presented in Appendix~\ref{oprep} (see Eqs.~\eqref{trans2}). The resultant Bose-Fermi analog of Hamiltonian \eqref{ham0} is considered in the next section.

We use also in Appendix~\ref{method1} the above described general procedure to construct representations of operators for one site in the unit cell. We only briefly consider in Appendix~\ref{method1} the resultant one-site formalism because the obtained representation is very similar to that derived in Ref.~\cite{chang} from other considerations and because self-consistent calculations are required in the one-site formalism to achieve sufficiently good quantitative agreement with previous numerical results for polaron properties.

\section{$t$--$J$ model on the square lattice through the bond-operator formalism}
\label{hafbi}

It is shown below that representation \eqref{trans2} allows to find all observable quantities as series in powers of $1/n$ using conventional diagrammatic technique. For this purpose, it is convenient to make all constant terms in the Hamiltonian of the order of $(1/n)^{-2}$, terms linear in Bose and Fermi operators of $(1/n)^{-3/2}$ order, bilinear terms of the order of $(1/n)^{-1}$, etc. Notice that some terms in the Hamiltonian satisfy this rule automatically (e.g., the term in Eq.~\eqref{ham0} containing ${\bf s}_p{\bf s}_j$) whereas some terms need to be multiplied by some power of $n$ (e.g., the formal renormalization $t \mapsto nt$ is needed in the first term in Eq.~\eqref{ham0}).
\footnote{Notice that we put factors $n^2$ and $n$ before the first and the last three terms in Eq.~\eqref{ss}, respectively, in order to make these terms of proper orders in $1/n$ in the Bose-analog of the spin Hamiltonian.} 
In this case, we use that only $n=1$ has the physical meaning. As a result, $n$ plays in the proposed formalism the role very much similar to the spin value $S$ in the Holstein-Primakoff transformation with the only difference that only $n=1$ has the physical meaning. We set $t=1$ in this section.

\subsection{Hamiltonian transformation. Harmonic approximation.}

Substituting Eqs.~\eqref{trans2} to Hamiltonian \eqref{ham0} and expanding the square root in operator \eqref{proj}, one obtains 
\begin{equation}
\label{hbf}
	{\cal H} = {\cal E} + \sum_{i=1}^\infty{\cal H}_i,
\end{equation}
where $\cal E$ is the ground state energy and ${\cal H}_i$ stand for terms containing products of $i$ operators of creation and annihilation. In particular, we have
\begin{eqnarray}
\label{e0b}
\frac{{\cal E}}{N} &=& -n^2\frac{J}{8} (4\sin2\alpha + 3\cos4\alpha + 5), \\
\label{h1b}
\frac{{\cal H}_1}{\sqrt N} &=& n^{3/2} \frac{J}{4} 
(2 \cos 2\alpha - 3\sin4\alpha) 
\left(a_{3\bf0}^{}+a^\dagger_{3\bf0}\right),\\
\label{h2}
{\cal H}_2 &=& {\cal H}_2^{(m)} + {\cal H}_2^{(1h)} + {\cal H}_2^{(2h)}, \\
\label{h2m}
{\cal H}_2^{(m)} &=& \sum_{\bf k} \left(
A_{\bf k} \left( a_{1\bf k}^\dagger a_{1\bf k}^{} + a_{2\bf k}^\dagger a_{2\bf k}^{} \right) 
+ B_{\bf k} a_{1\bf k}^{} a_{2-\bf k}^{} + B_{\bf k}^* a_{1\bf k}^\dagger a_{2-\bf k}^\dagger 
+
C_{\bf k} a_{3\bf k}^\dagger a_{3\bf k}^{} 
+ \frac{D_{\bf k}}{2} \left( a_{3\bf k}^\dagger a_{3-\bf k}^\dagger + a_{3\bf k}^{} a_{3-\bf k}^{} \right) 
\right),\\
\label{h22f}
{\cal H}_2^{(2h)} &=& n\sum_{\bf k} \sum_{j=1}^4
\frac{J}{4} (2 \sin2\alpha + 3\cos4\alpha + 11)b_{j\bf k}^\dagger b_{j\bf k}^{},
\end{eqnarray}
where $N$ is the number of unit cells in the lattice, ${\cal H}_2^{(m)}$, ${\cal H}_2^{(1h)}$, and ${\cal H}_2^{(2h)}$ stand for bilinear terms containing magnetic, one-holon fermionic, and two-holon bosonic operators, respectively, we omit in Eq.~\eqref{e0b} the term $-n^2J/2$ stemming from the last term in Eq.~\eqref{ham0},
\begin{eqnarray}
\label{coefb}
A_{\bf k} &=& 
n\frac{J}{4} (3 \cos4\alpha - 2\sin2\alpha ({\rm Re}\nu_{\bf k}-1) + 5),\nonumber\\
B_{\bf k} &=& n \frac{J}{2} \left(\nu_{\bf k} \sin^2\alpha + \nu_{\bf k}^* \cos^2\alpha\right),\nonumber\\
C_{\bf k} &=& 
n\frac{J}{2} (2\sin2\alpha + 6\cos^22\alpha - {\rm Re}\nu_{\bf k}\sin^22\alpha),\\
D_{\bf k} &=& -n \frac{J}{4} {\rm Re}\nu_{\bf k} \sin ^22\alpha,\nonumber\\
\nu_{\bf k} &=& e^{i (k_1+k_2)}+e^{i k_1}+e^{i k_2},\nonumber
\end{eqnarray}
and $k_p=({\bf k}\mbox{\boldmath $\mathfrak v$}_p)$ ($p=1,2$) are components of $\bf k$ (see Fig.~\ref{system}(b)). 

Minimization of ${\cal E}$ given by Eq.~\eqref{e0b} gives the following value $\alpha_0$ of $\alpha$ in the leading order in $1/n$: $\sin2\alpha_0 = 1/3$.
At $\alpha=\alpha_0$, linear terms \eqref{h1b} vanish in the Hamiltonian and one obtains in the harmonic approximation $\frac{\cal E}{N} = -n^213/12$ and the staggered magnetization
$M=\langle S_{1j}^z \rangle=-\langle S_{2j}^z \rangle=\frac n2\cos2\alpha_0=n\sqrt2/3\approx0.47n$.

${\cal H}_2^{(1h)}$ in Eq.~\eqref{h2} contains 40 terms of the form $r^\dagger_{i\bf k}f^{}_{j\bf k}$, where $r,f=e,d$, which we do not present here due to their cumbersomeness. However, ${\cal H}_2^{(1h)}$ has a simple form 
\begin{subequations}
\label{h21f}
\begin{eqnarray}
\label{h21fd}
{\cal H}_2^{(1h)} &=& \sum_{\bf k}\left(
\epsilon_0^{(d1)} \left( \tilde d_{1\bf k}^\dagger \tilde d_{1\bf k}^{} + \tilde d_{2\bf k}^\dagger \tilde d_{2\bf k}^{} \right)
+
\epsilon_0^{(d2)} \left( \tilde e_{1\bf k}^\dagger \tilde e_{1\bf k}^{} + \tilde e_{2\bf k}^\dagger \tilde e_{2\bf k}^{} \right)\right.\\
\label{h21fv}
&&\left.{}
+
E_{\bf k} \left( \tilde d_{3\bf k}^\dagger \tilde d_{3\bf k}^{} + \tilde d_{4\bf k}^\dagger \tilde d_{4\bf k}^{} \right)
+
H_{\bf k} \left( \tilde e_{3\bf k}^\dagger \tilde e_{3\bf k}^{} + \tilde e_{4\bf k}^\dagger \tilde e_{4\bf k}^{} \right)
+
I_{\bf k} \left( \tilde d_{3\bf k}^\dagger \tilde e_{3\bf k}^{} + \tilde e_{4\bf k}^\dagger \tilde d_{4\bf k}^{} \right)
+
I^*_{\bf k} \left( \tilde e_{3\bf k}^\dagger \tilde d_{3\bf k}^{} + \tilde d_{4\bf k}^\dagger \tilde e_{4\bf k}^{} \right)
\right)
\end{eqnarray}
\end{subequations}
when written in terms of new fermionic operators $\tilde e_j,\tilde d_j$ related with the initial fermions by a unitary transformation 
\begin{equation}
\label{u}
	(\tilde e_1,\tilde e_2,\tilde e_3,\tilde e_4,\tilde d_1,\tilde d_2,\tilde d_3,\tilde d_4)
	=
	(e_1,e_2,e_3,e_4,d_1,d_2,d_3,d_4)U,
\end{equation} 
where $U$ is a unitary matrix with real coefficients which depend on $J$ and $\alpha$ and do not depend on $\bf k$. We failed to obtain a compact analytical expression for $U$ at arbitrary $J$ and $\alpha$. In particular, coefficients in Eq.~\eqref{h21f} have the following form at $J=0.3$ and $\alpha=\alpha_0$: 
\begin{eqnarray}
\label{coeff}
\epsilon_0^{(d1)} &\approx& n0.613,\nonumber\\
\epsilon_0^{(d2)} &\approx& n1.037,\nonumber\\
E_{\bf k} &\approx& n (1.828 - 0.153 {\rm Re}\nu_{\bf k} ),\\
H_{\bf k} &\approx& n (-0.178 + 0.153 {\rm Re}\nu_{\bf k} ),\nonumber\\
I_{\bf k} &\approx& n (-0.195 -0.051\nu_{\bf k} + 0.116\nu^*_{\bf k}),\nonumber
\end{eqnarray}
where $\nu_{\bf k}$ is defined in Eq.~\eqref{coefb}. 

Notice that one would have come to Eq.~\eqref{h21f} for ${\cal H}_2^{(1h)}$ if one had used basis functions for Fermi operators which are related with those presented in Table~\ref{statestab} by the unitary transformation $U$ from Eq.~\eqref{u}. It is convenient in the subsequent consideration to switch from operators $d_j,e_j$ to $\tilde d_j,\tilde e_j$ related by Eq.~\eqref{u} in which $U$ is found at $\alpha=\alpha_0$ (in this case, ${\cal H}_2^{(1h)}$ has the form \eqref{h21f} in the leading order in $1/n$).

\subsection{Quasiparticles Green's functions and diagrammatic technique} 

\begin{figure}
\includegraphics[scale=0.7]{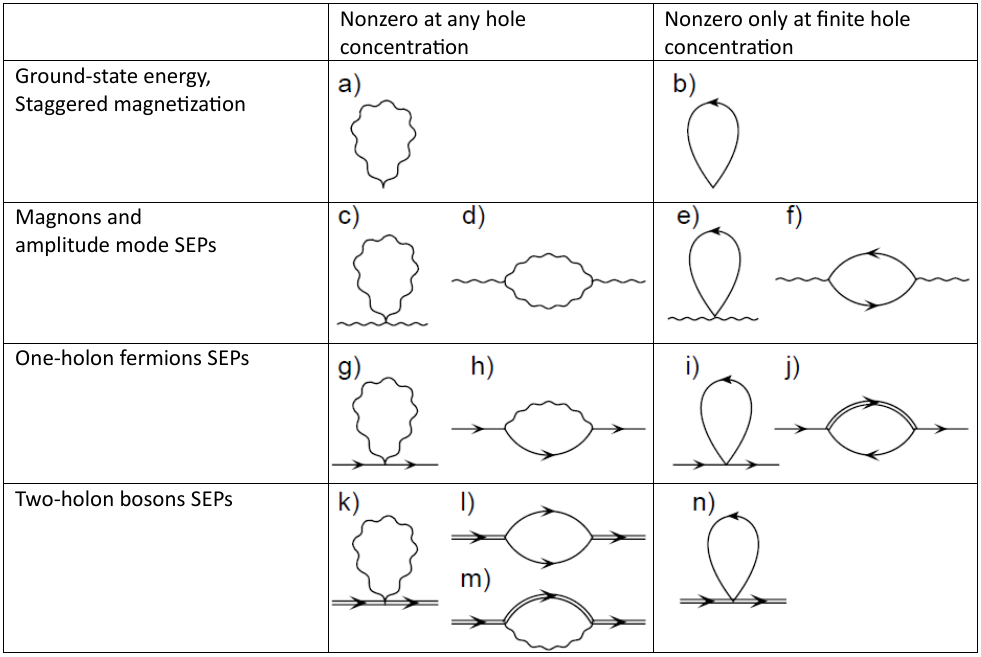}
\caption{Diagrams giving corrections of the first order in $1/n$ to the ground state energy, the staggered magnetization, and to self-energy parts (SEPs) of magnons, the amplitude mode, one-holon fermions, and two-holon bosons. Wavy lines denote magnon and/or Higgs mode. Straight and double straight lines stand for Green's functions of one-holon fermions and two-holon bosons, respectively.
\label{diag}}
\end{figure}

First $1/n$ corrections to observable quantities can be found by the conventional diagrammatic technique. As in Refs.~\cite{inem,syromyat,ibot}, we use an approach not involving Bogoliubov transformations which operates with Green's functions of the type
$G_{QF}(\omega,{\bf k}) = \langle Q_{\bf k}, F_{-\bf k} \rangle_\omega
= -i\frac{1}{N}\sum_{j,p} e^{-i{\bf k}{\bf R}_{jp}}
\int dt e^{i\omega t}	
\langle T Q_j(t)F_p(0)\rangle$, 
where $Q,F=a^{}_{1,2},a_{1,2}^\dagger$ for magnons, $Q,F=a^{}_3,a_3^\dagger$ for the Higgs mode, $Q=\tilde d_j,\tilde e_j$, $F=\tilde d_j^\dagger,\tilde e_j^\dagger$ for fermions, and $Q=b_j$, $F=b_j^\dagger$ for two-holon bosons. Then, one deals with sets of Dyson equations for Green's functions within this approach. Such a technique is more compact and, thus, more convenient for cumbersome calculations. 

In particular, one has for magnon Green's functions
\begin{equation}
\label{eqfunc}
\begin{array}{l}
G_{11}(k) = G_{11}^{(0)}(k) + G_{11}^{(0)}(k)\Sigma_{11}(k)G_{11}(k) + G_{11}^{(0)}(k) [B^*_{\bf k} + {\overline \Pi_{12}}(k)] G_{21}(k),\\
G_{21}(k) = {\overline G}_{22}^{(0)}(k) {\overline \Sigma}_{22}(k)G_{21}(k) + {\overline G}_{22}^{(0)}(k) [B_{\bf k} + \Pi_{21}(k) ]G_{11}(k),
\end{array}
\end{equation}
where $k=(\omega,{\bf k})$, 
$G_{11}(k)=\langle a^{}_{1\bf k}, a^\dagger_{1\bf k} \rangle_\omega$, 
$G_{21}(k)=\langle a^\dagger_{2-\bf k}, a^\dagger_{1\bf k} \rangle_\omega$, 
$G_{11}^{(0)}(k) = {\overline G}_{22}^{(0)}(-k) = (\omega - A_{\bf k})^{-1}$, and $\Sigma_{11}(k)$, ${\overline \Sigma}_{22}(k)$, $\Pi_{21}(k)$, ${\overline \Pi}_{12}(k)$ are self-energy parts, and $A_{\bf k}$ and $B_{\bf k}$ are given by Eq.~\eqref{coefb}. One obtains solving Eqs.~(\ref{eqfunc})
\begin{eqnarray}
G_{11}(k) &=& \frac{\omega + A_{\bf k} + {\overline \Sigma}_{22}(k)}{{\cal D}(k)},\nonumber\\
\label{gf}
G_{21}(k) &=& -\frac{B_{\bf k} + \Pi_{21}(k)}{{\cal D}(k)},
\end{eqnarray}
where
\begin{eqnarray}
\label{d}
{\cal D}(k) &=& \omega^2 - \left(\epsilon^{(m)}_{0\bf k}\right)^2 - \Omega(k),\\
\label{specm0}
\epsilon^{(m)}_{0\bf k} &=& \sqrt{A_{\bf k}^2 - \left|B_{\bf k}\right|^2},\\
\label{o}
\Omega(k) &=& A_{\bf k}(\Sigma_{11} + \overline{\Sigma}_{22}) 
- B_{\bf k}{\overline \Pi}_{12} - B^*_{\bf k}\Pi_{21} 
+ \omega (\Sigma_{11} - \overline{\Sigma}_{22}) 
- \Pi_{21}{\overline \Pi}_{12} + \Sigma_{11} \overline{\Sigma}_{22},
\end{eqnarray}
$\epsilon^{(m)}_{0\bf k}$ is the bare magnon spectrum and $\Omega(k)$ describes its renormalization. Remaining Green's functions of magnons, the Higgs mode, and fermions can be obtained similarly. The bare spectrum of the amplitude mode has the form
\begin{equation}
\label{speca0}
\epsilon_{0\bf k}^{(a)} = \sqrt{C_{\bf k}^2 - D_{\bf k}^2},
\end{equation}
where $C_{\bf k}$ and $D_{\bf k}$ are given by Eq.~\eqref{coefb}. Eq.~\eqref{h21f} gives four doubly degenerate fermionic modes with spectra $\epsilon_0^{(d1)}$, $\epsilon_0^{(d2)}$, and
\begin{equation}
\label{specf0}
	\epsilon_{0\bf k}^{(v\pm)} = \frac12 \left( 
	E_{\bf k} + H_{\bf k} \pm \sqrt{ (E_{\bf k} - H_{\bf k})^2 +4 |I_{\bf k}|^2 }
	\right).
\end{equation}

${\cal H}_3$ and ${\cal H}_4$ terms in Hamiltonian \eqref{hbf} lead to diagrams of the first order in $1/n$ for self-energy parts shown in Fig.~\ref{diag} (as usual, they should be calculated using bare Green's functions in which all self-energy parts are discarded). Besides, as soon as coefficients in the Hamiltonian depend on $\alpha$, the renormalization of $\alpha$ also contributes to the renormalization of observables. By making all possible couplings of operators $a$ in ${\cal H}_3$ (taken at $\alpha=\alpha_0$), one derives at zero hole concentration the first-order correction to ${\cal H}_1$ and obtains the correction to $\alpha_0$ from the requirement that ${\cal H}_1$ should vanish. 

Notice that the operator of the number of holes has the form in our consideration
\begin{equation}
\label{caln}
	{\cal N} = \sum_{i=1}^4 \left(\tilde d_i^\dagger \tilde d_i + \tilde e_i^\dagger \tilde e_i + 2 b_i^\dagger b_i \right).
\end{equation}
Then, the chemical potential of two-holon bosons is two time greater than the chemical potential of fermions at finite concentration of holes. We consider in the present paper only cases of zero, one, and two holes. Then, the chemical potential of all quasiparticles is zero in subsequent calculations and Green's functions of $\tilde d$, $\tilde e$, and $b$ operators are analytic in the upper half-plane so that diagrams shown in Fig.~\ref{diag} in the right column give zero. Spectra of one- and two-holon quasiparticles obtained in this case by the diagrammatic technique are counted from the energy of the N\'eel ground state of the system without holes.

\begin{figure}
\includegraphics[scale=0.1]{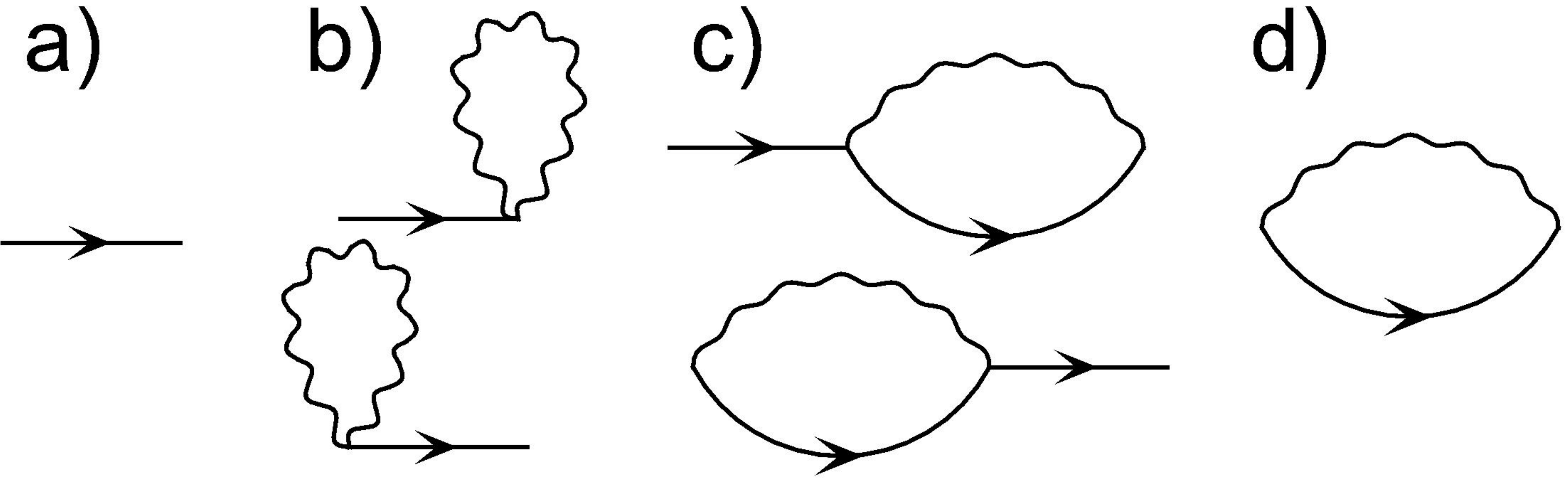}
\caption{Diagrams giving corrections of the first order in $1/n$ to electron Green's function \eqref{egf} at zero concentration of holes. As in Fig.~\ref{diag}, wavy lines denote magnon and/or Higgs mode and straight lines stand for fermions Green's functions.
\label{chifig}}
\end{figure}

It is seen from Eqs.~\eqref{cop}, \eqref{cu2}--\eqref{cd2} that the electron Green's function
\begin{equation}
\label{egf}
	\langle c^\dagger_{\bf k\sigma}, c^{}_{\bf k\sigma} \rangle_\omega
= -i\int dt e^{i\omega t}	
\langle T c^\dagger_{\bf k\sigma}(t)c^{}_{\bf k\sigma}(0)\rangle
\end{equation}
is related with Green's functions $\langle Q^{}_{\bf k}, F^\dagger_{\bf k} \rangle_\omega$ of fermionic operators $Q,F=d_j,e_j$. Then, fermionic quasiparticles appearing in our theory produce poles in electronic Green's function \eqref{egf}. Importantly, terms in $c_{\bf k\sigma}$ which are linear in Fermi-operators contain only $\tilde d_{3\bf k}$, $\tilde d_{4\bf k}$, $\tilde e_{3\bf k}$, and $\tilde e_{4\bf k}$ operators which are introduced in Eq.~\eqref{u}. As soon as observable quantities related with mobile electrons are expressed via electronic Green's functions built on $c_{\bf k\sigma}$ and $c_{\bf k\sigma}^\dagger$, we call fermions as follows below:
\begin{eqnarray}
\tilde d_{1\bf k},\tilde d_{2\bf k},\tilde e_{1\bf k},\tilde e_{2\bf k} &-& \mbox{dark (invisible) fermions,} 
\nonumber\\
\tilde d_{3\bf k},\tilde d_{4\bf k},\tilde e_{3\bf k},\tilde e_{4\bf k} &-& \mbox{visible fermions.}
\nonumber
\end{eqnarray}
Their bare spectra are determined by Eqs.~\eqref{h21fd} and \eqref{h21fv}, correspondingly (see also Eqs.~\eqref{coeff} and \eqref{specf0}). Diagrams are shown in Fig.~\ref{chifig} which give corrections of the first order in $1/n$ to electron Green's function \eqref{egf} at zero concentration of holes. Diagrams presented in Fig.~\ref{chifig}(a)--\ref{chifig}(c) contain fermion Green's functions and thus obey sharp anomalies produced by quasiparticles whereas diagram shown in Fig.~\ref{chifig}(d) contributes to an incoherent background.

\subsection{Static characteristics in the first order in $1/n$} 
\label{statics}

Corrections to the ground state energy $\cal E$ and to the staggered magnetization $M$ come from the $\alpha$ renormalization and from all possible couplings of Bose operators in ${\cal H}_2$ and in bilinear terms in Eqs.~\eqref{s1z2} and \eqref{s2z2}, respectively (see the diagram in Fig.~\ref{diag}(a)). One obtains after simple calculations
\begin{eqnarray}
\alpha &=& \frac12 \arcsin\frac13 + 0.0982\frac1n,\\
\label{eval}
\frac{\cal E}{N} &=& -\frac{13}{12}n^2 - 0.2255n,\\
\label{mval}
M &=& \langle S_{1j}^z \rangle=-\langle S_{2j}^z \rangle = \frac{\sqrt2}{3}n - 0.1498.
\end{eqnarray}
We have at $n=1$ from Eqs.~\eqref{eval} and \eqref{mval} $M\approx0.32$ and $\frac{\cal E}{2N}\approx-0.654$ which are very close to respective values of $\approx0.3$ and $\approx -0.667$ found before by many other methods (see, e.g., Ref.~\cite{monous}).

It should be noted that the proposed BOT breaks the equivalence of bonds inside unit cells and between adjacent unit cells in the first few orders in $1/n$. This inequivalence is certainly an artefact of the approach caused by the truncation of the $1/n$ series. However this artefact is very small in the first order in $1/n$. In particular, we find that $\langle {\bf S}_i {\bf S}_j \rangle$ has the form 
$ -\frac{5}{12} n^2 + 0.073 n $ and $ -\frac29 n^2 - 0.099 n $ 
if nearest neighbor sites $i$ and $j$ belong to the same and to adjacent unit cells, respectively. These results read at $n=1$ as $-0.344$ and $-0.322$, correspondingly, which differ only 6\%.  

\subsection{Spectra of purely magnetic quasiparticles}

\begin{figure}
\includegraphics[scale=0.4]{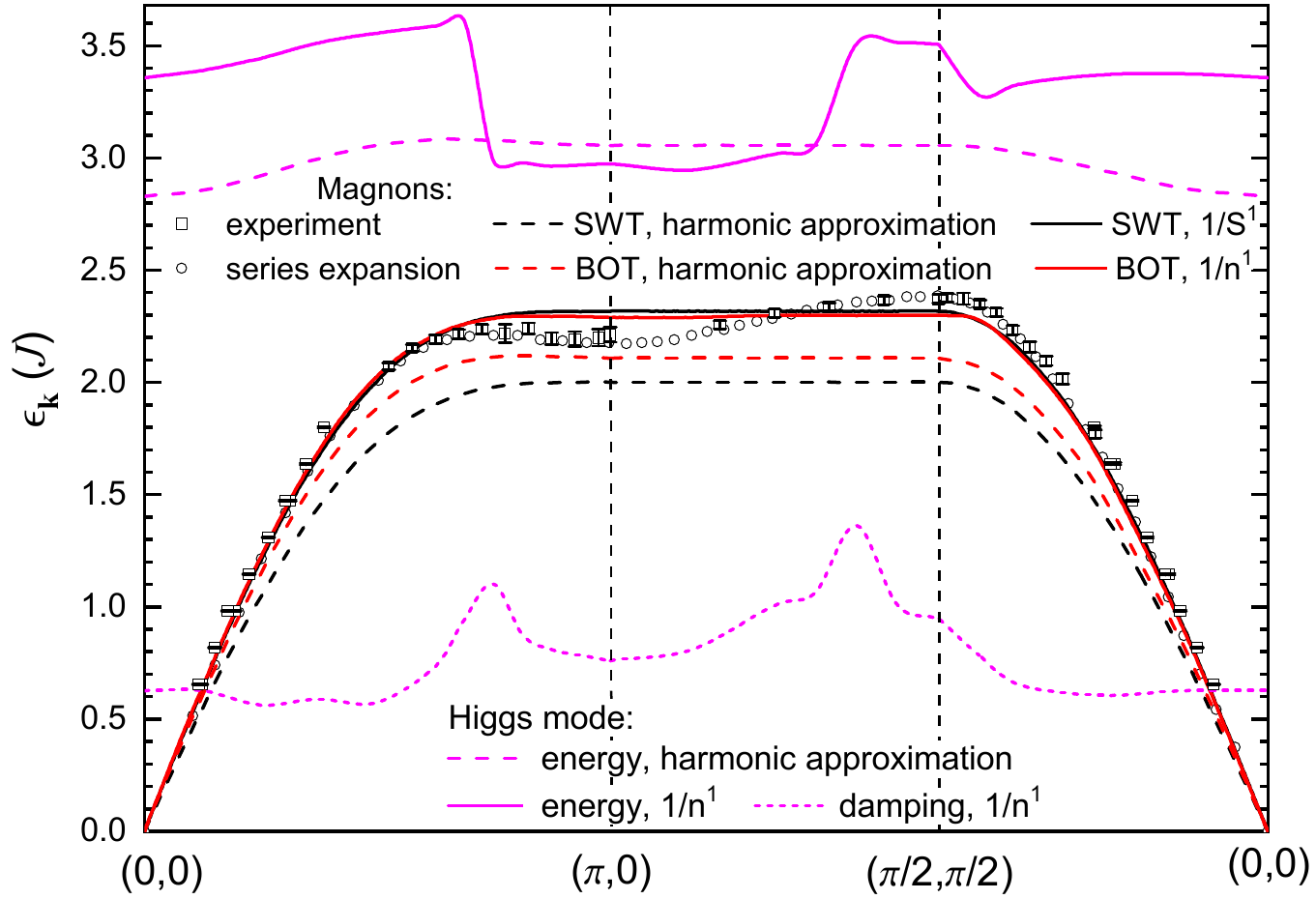}
\caption{Spectra in units of $J$ of magnetic elementary excitations, magnons and the amplitude (Higgs) mode, in $t$--$J$ models at zero concentration of holes found using the proposed bond-operator technique (BOT). Also shown are magnon spectra obtained by series expansion around the Ising limit \cite{ser}, within the spin-wave theory (SWT), and neutron scattering experiment in a molecular compound abbreviated as CFTD \cite{chris1,piazza} which is described well by spin-$\frac12$ square-lattice Heisenberg antiferromagnet. Points in the reciprocal space are denoted according to Fig.~\ref{system}(b).
\label{magnons}}
\end{figure}

Spectra of magnons and the amplitude mode found in the harmonic approximation (bare spectra given by Eqs.~\eqref{specm0} and \eqref{speca0}) and in the first order in $1/n$ are shown in Fig.~\ref{magnons}. It is seen that similar to results of the spin-wave theory obtained in the first order in $1/S$, magnon spectrum obtained in the first order in $1/n$ agrees quantitatively with previous numerical and experimental findings. The Higgs mode lies above magnons and acquires a large damping in the first order in $1/n$ due to the decay into two magnons described by the diagram shown in Fig.~\ref{diag}(d).

\subsection{Spectra of fermionic quasiparticles}

\begin{figure}
\includegraphics[scale=0.4]{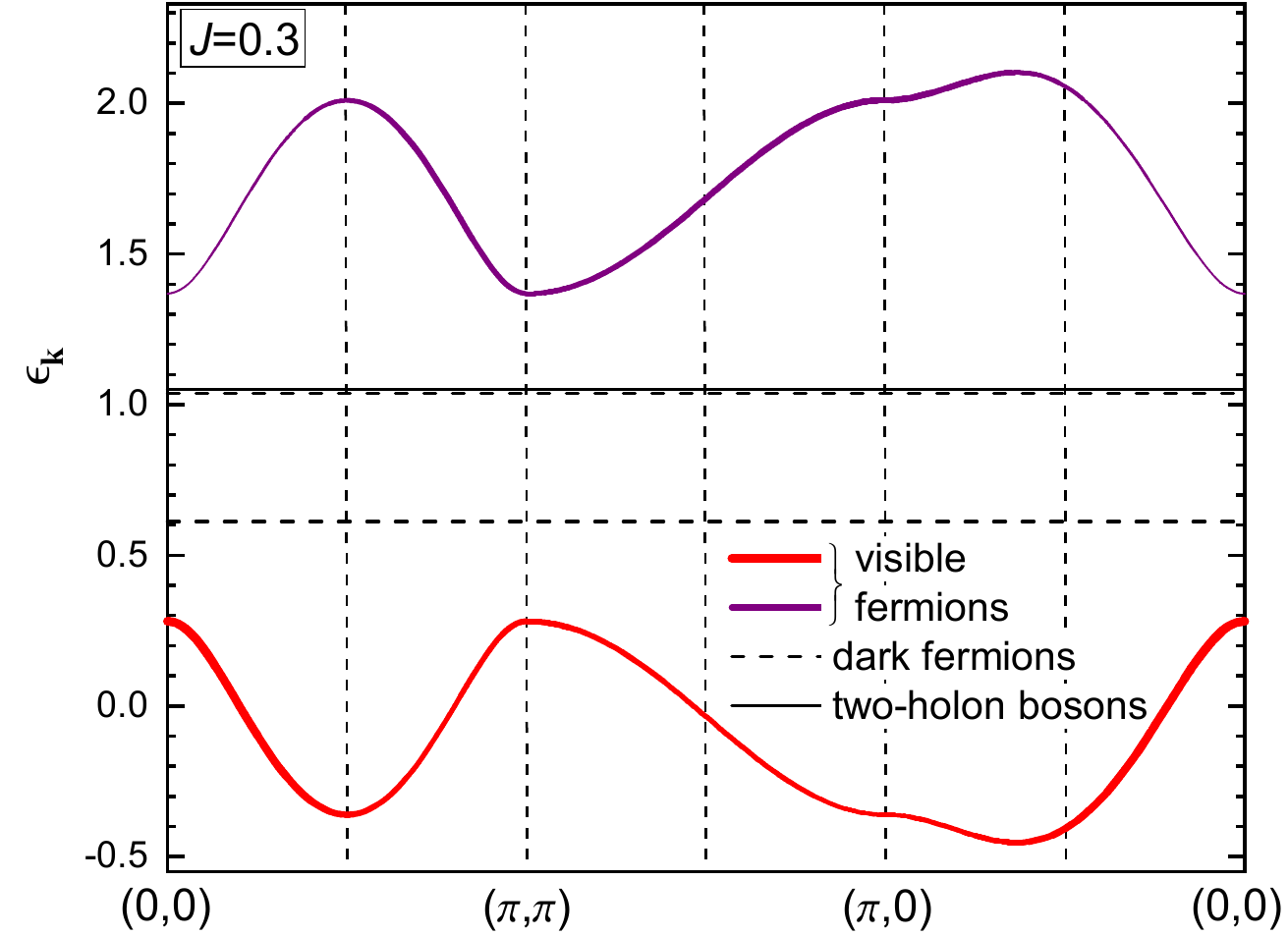}
\caption{
Spectra of one- and two-holon elementary excitations obtained in the harmonic approximation of the BOT in the $t$--$J$ model at $J=0.3$. Thickness of a solid line is proportional to the spectral weight of visible fermions in the electron Green's function \eqref{egf}. 
\label{haall}}
\end{figure}

Fig.~\ref{haall} shows spectra of one-holon elementary excitations obtained in the harmonic approximation of the BOT in the $t$--$J$ model at $J=0.3$. There are two dark modes which are dispersionless in the harmonic approximation and two visible modes whose spectral weight in the electronic Green's function \eqref{egf} is indicated by the line width.

\begin{figure}
\includegraphics[scale=0.4]{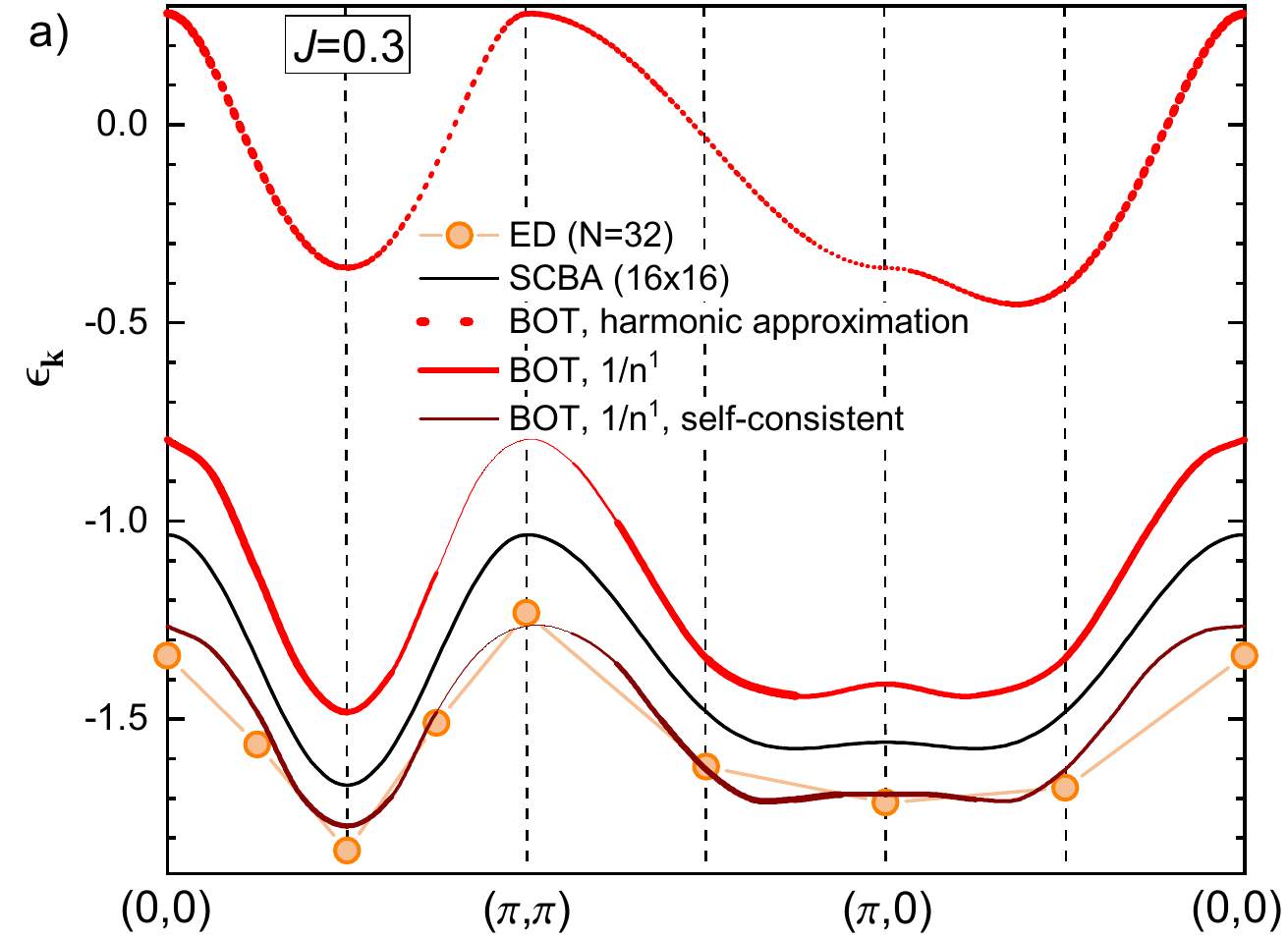}
\includegraphics[scale=0.4]{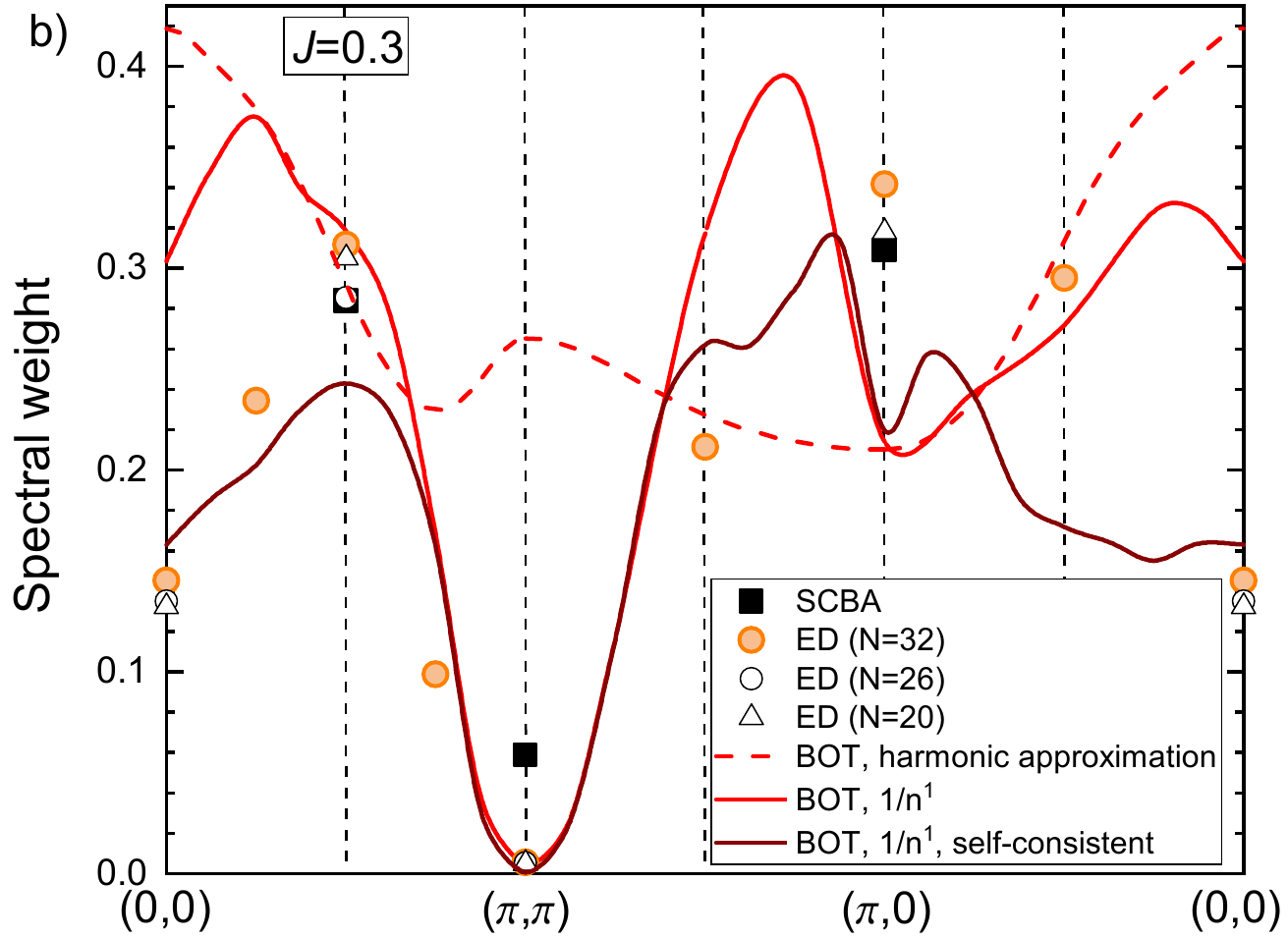}
\caption{
a) Spectra of the electron Green's function $\sum_\sigma\langle c^\dagger_{\bf k\sigma}, c^{}_{\bf k\sigma} \rangle_\omega$ given by Eq.~\eqref{egf} found at $J=0.3$ using exact diagonalization (ED) of a cluster with 32 sites (data are taken from Fig.~3 of Ref.~\cite{ed1}), using the formula 
$x_1 + x_2(\cos k_x + \cos k_y)^2 + x_3(\cos 2k_x + \cos 2k_y)$ 
with coefficients $x_{1,2,3}$ obtained in the self-consistent Born approximation (SCBA) on $16\times16$ cluster (see Table~II in Ref.~\cite{scba1}), and in the BOT within the harmonic approximation and in the first order in $1/n$. Thickness of a solid line in the BOT data is proportional to the spectral weight. Also shown are results found self-consistently in the first order in $1/n$ as it is described in the text. b) Spectral weights of modes shown in panel a). ED data for $N=26$ and 20 are taken from Ref.~\cite{ed2}.
\label{spec03}}
\end{figure}

All fermionic modes are renormalized considerably in the first order in $1/n$. The major renormalization comes from the loop diagrams shown in Fig.~\ref{diag}(h). The spectrum of the lower fermionic mode (magnetic polaron) and its spectral weight found in the first order in $1/n$ at $J=0.3$ are shown in Fig.~\ref{spec03}. It is seen that our findings are in a good agreement with previous results obtained in the self-consistent Born approximation (SCBA) and by exact diagonalization of finite clusters. In particular, the spectrum form and its minimum at ${\bf k}=({\pi/2,\pi/2})$ are reproduced by the BOT quite accurately. Renormalized spectra of other fermionic quasiparticles still lie significantly above the lower branch and we do not consider them in this subsection.

The spectral weight found in the first order in $1/n$ and presented in Fig.~\ref{spec03}(b) is in a good quantitative agreement with previous findings except for the vicinity of $\bf k=0$. We point out that the contribution of diagrams shown in Figs.~\ref{chifig}(b) and \ref{chifig}(c) to the spectral weight in Fig.~\ref{spec03}(b) is small in the entire Brillouin zone except for the vicinity of the point ${\bf k}=({\pi,\pi})$. The diagram presented in Fig.~\ref{chifig}(a) alone gives the spectral weight at ${\bf k}=({\pi,\pi})$ which is very close to the result of the SCBA (see Fig.~\ref{spec03}(b)). Taking into account diagrams shown in Figs.~\ref{chifig}(b) and \ref{chifig}(c) which were ignored in previous calculations within the SCBA \cite{scba1} brings our findings near ${\bf k}=({\pi,\pi})$ into a quantitative agreement with the exact diagonalization results.

We find also fermionic spectra self-consistently by iterative calculations of diagrams arising in the first order in $1/n$ (i.e., at the first step, we calculate self-energy parts using bare Green's functions, at the second step, we find self-energy parts using Green's functions obtained at the first step, etc.). The convergence of this procedure occurs after 3--5 steps depending on model parameters (the general trend is that the smaller $J$ value, the larger number of steps is required for the convergence). 
For $J=0.3$, corresponding results are also shown in Fig.~\ref{spec03}. It is seen that the obtained polaron energy reproduces well numerical results. The overall agreement in the case of the spectral weight is also becomes better.

\begin{figure}
\includegraphics[scale=0.4]{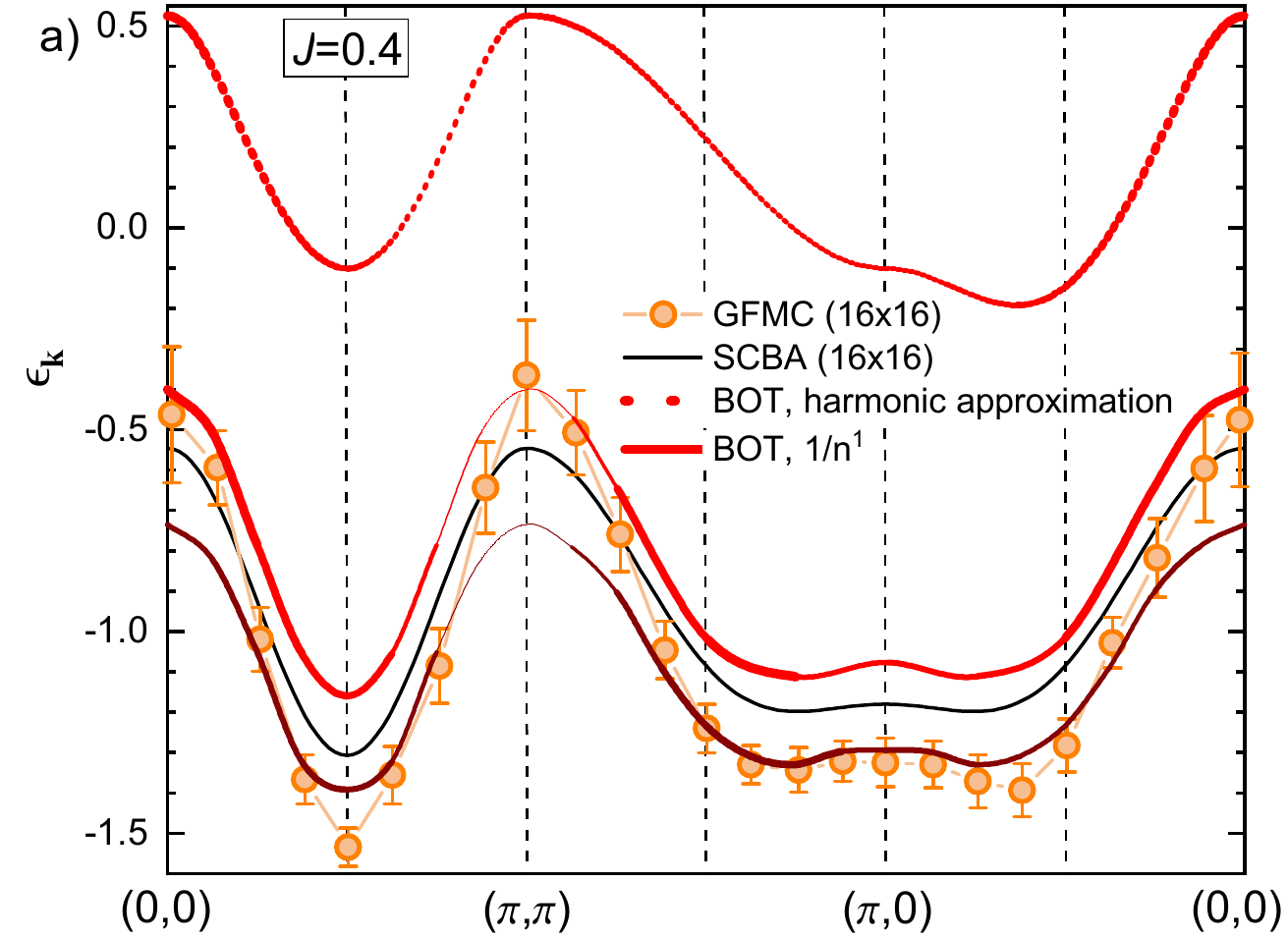}
\includegraphics[scale=0.4]{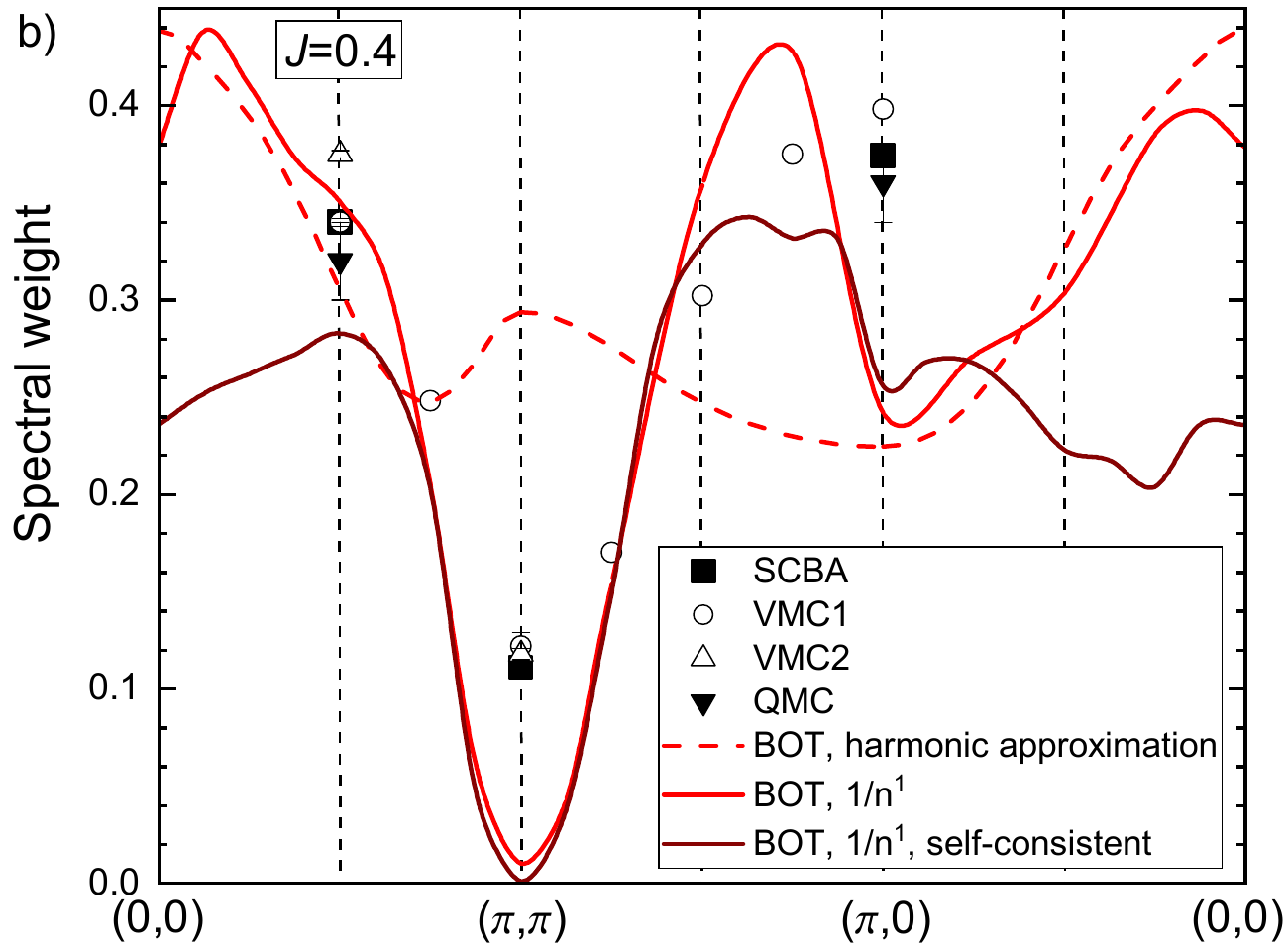}
\caption{
Same as Fig.~\ref{spec03} but for $J=0.4$. Also presented are data of Green's function Monte Carlo (GFMC) simulations (taken from Fig.~1 of Ref.~\cite{gf2}), variational Monte Carlo calculations using different trial wave functions (VMC1 \cite{vmc2} and VMC2 \cite{vmc}), and Quantum Monte Carlo (QMC) simulations \cite{qmc}.
\label{spec04}}
\end{figure}

The agreement between the BOT and previous results are also good at $J=0.4$ as it is seen from Fig.~\ref{spec04}, where numerical data are shown which were obtained by another set of numerical techniques. The situation with the spectral weight near ${\bf k}=({\pi,\pi})$ is similar to that discussed above for $J=0.3$. 

The energy of the magnetic polaron and the spectral weight in the spectrum minimum (i.e., at ${\bf k}={\bf k}_m=({\pi/2,\pi/2})$) are in a good quantitative agreement with previous numerical findings at $J=0.1-1.0$ as it is seen from Figs.~\ref{ez}(a) and \ref{ez}(b). Our results for the polaron band width $W$ presented in Fig.~\ref{ez}(c) also reproduce well the most reliable numerical results and SCBA findings.

\begin{figure}
\includegraphics[scale=0.4]{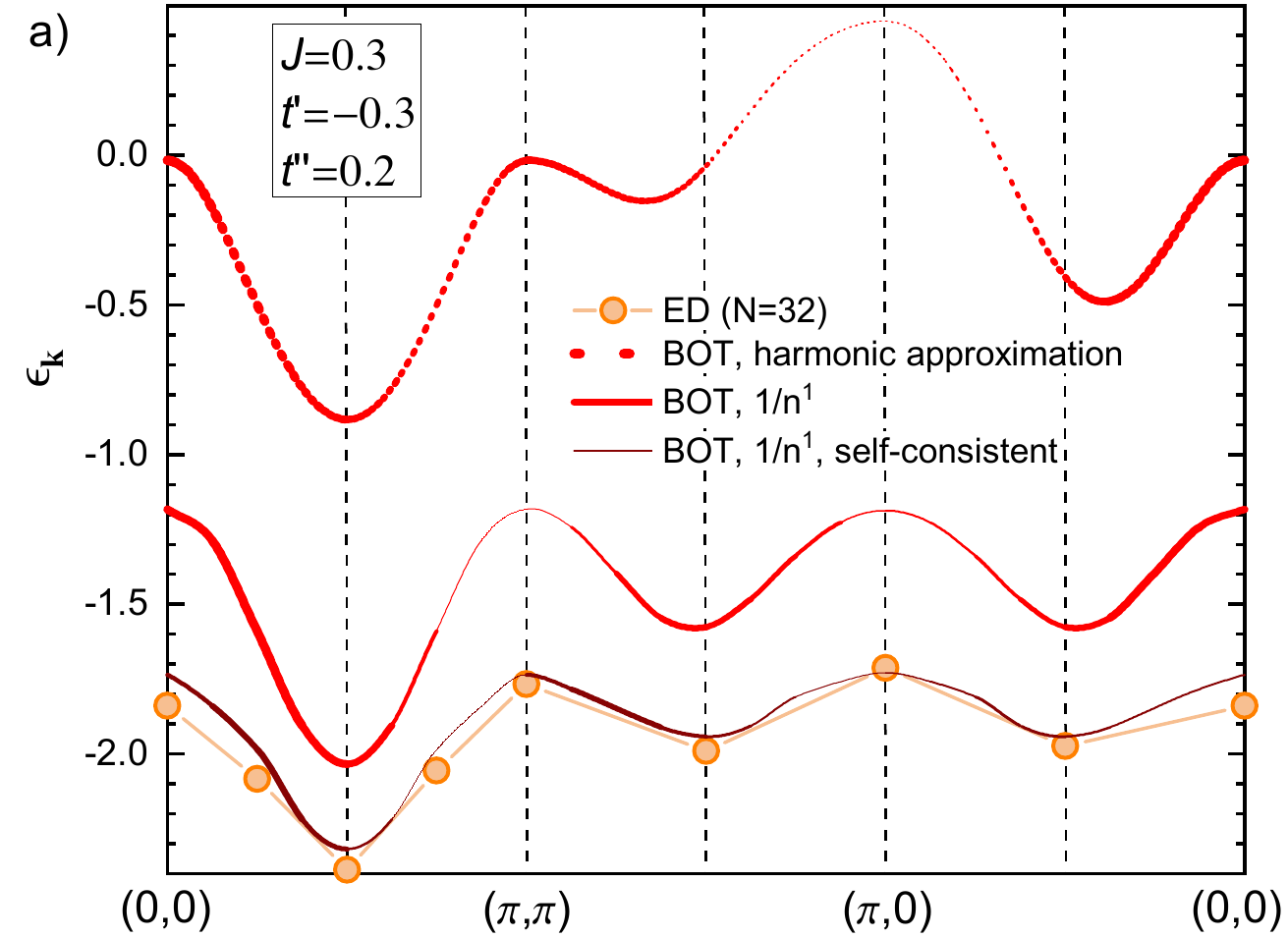}
\includegraphics[scale=0.4]{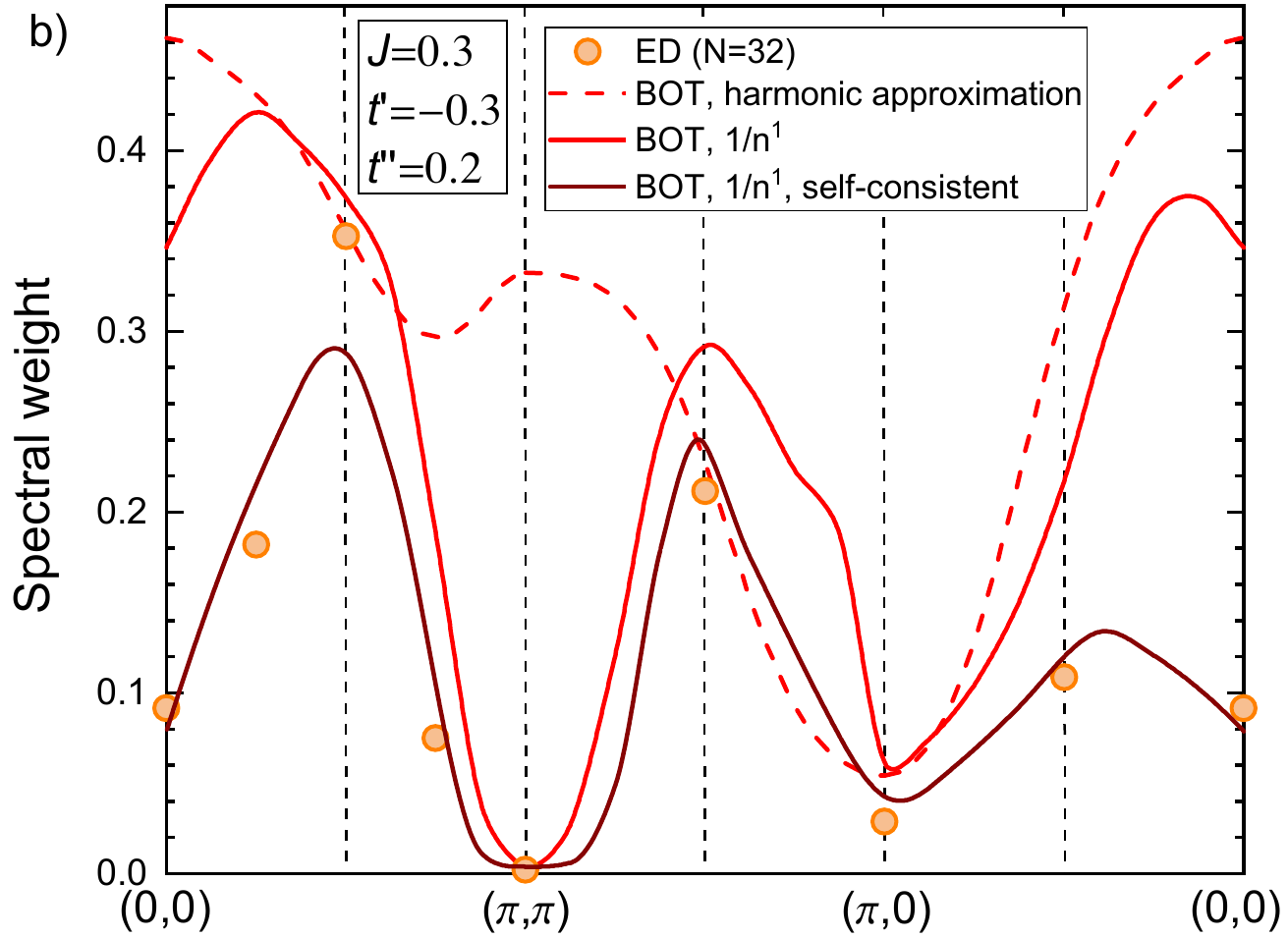}
\caption{
Same as Fig.~\ref{spec03} but for $J=0.3$, $t'=-0.3$, and $t''=0.2$. Exact diagonalization (ED) results of a cluster with $N=32$ sites are taken from Ref.~\cite{leung2}.
\label{spectttp}}
\end{figure}

The agreement between the BOT and previous numerical results remains also in extended $t$--$J$ models in which hoping between the second- and the third-nearest neighbor sites $t'$ and $t''$ are included. Fig.~\ref{spectttp} illustrates this for the set of parameters $J=0.3$, $t'=-0.3$, and $t''=0.2$ which was introduced in Ref.~\cite{leung2} to describe ARPES data in $\rm Sr_2CuO_2Cl_2$.

\subsection{Spectra of two-holon bosons. Bound states of two holes.}

It is seen from Eq.~\eqref{h22f} that there are four degenerate branches of two-holon bosons in the harmonic approximation who lie above the lower fermionic branch as Fig.~\ref{haall} illustrates for $J=0.3$. This degeneracy disappears in the first order in $1/n$ and the lower branch is renormalized downward considerably and much more significantly than fermionic spectra considered above. The main renormalization comes from the diagram shown in Fig.~\ref{diag}(l) which describes a decay of a two-holon boson into two fermions. Notice that the diagram in Fig.~\ref{diag}(m) describes the interaction of the bosons with the Higgs mode which gives a very small contribution. The Hartree-Fock diagram in Fig.~\ref{diag}(k) produces moderate positive corrections. 

Due to the large renormalization, one has to go beyond the first order in $1/n$ in order to achieve a reasonable agreement with previous numerical findings. For this purpose, we calculate diagrams in Fig.~\ref{diag}(l) using fermionic Green's functions found self-consistently as it is described above. We find that the lower bosonic branch acquires a considerable damping in the whole Brillouin zone due to the decay into two fermions for $J=0.1-0.8$. However at $J\agt0.8$ the boson spectrum $\epsilon_{b\bf k}$ is well defined around $\bf k=0$, where it lies below the two-fermion continuum. At an infinitesimal concentration of holes, the chemical potential $\mu$ of fermions should be introduced which is slightly larger than the polaron energy $\epsilon_{{\bf k}={\bf k}_m}$ at its minimum. As it is noted above, the chemical potential of two-holon bosons is $2\mu$ (see Eq.~\eqref{caln}). Then, our finding that $\epsilon_{b\bf k}$ lies below the two-fermion continuum near $\bf k=0$ at $J\agt0.8$ reads as $\epsilon_{b\bf k}-2\mu<0$ at $\bf k=0$ that signifies a condensation of the two-holon bosons and a superconducting state at a finite hole concentration.

\begin{figure}
\includegraphics[scale=0.4]{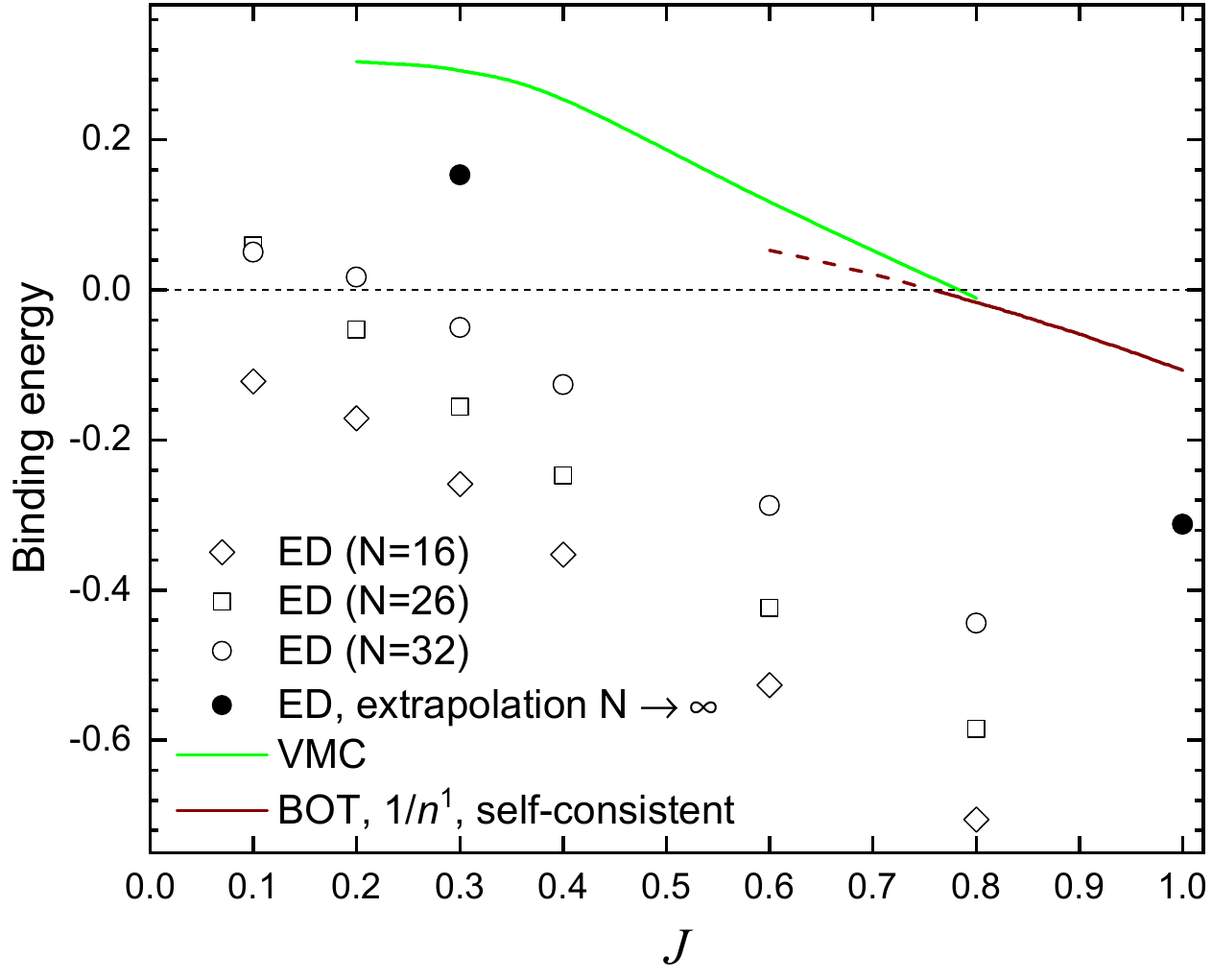}
\caption{
Binding energy of two holes obtained in exact diagonalization (ED) and valence Monte Carlo (VMC) studies which are taken from Figs.~6 of Ref.~\cite{pairstj} and Ref.~\cite{pairsvmc}, respectively. Also shown are results of our calculation of $\epsilon_{b\bf k}-2\epsilon_{{\bf k}_m}$ at $\bf k=0$, where $\epsilon_{b\bf k}$ is the spectrum of the lower two-holon branch and $\epsilon_{{\bf k}_m}$ is the polaron energy at its minimum at ${\bf k}={\bf k}_m$. The dashed part of the line corresponds to $\epsilon_{b\bf 0}-2\epsilon_{{\bf k}_m}>0$ and denotes that $\epsilon_{b\bf 0}$ has a finite damping.
\label{pairs}}
\end{figure}

To verify our conclusion, we compare in Fig.~\ref{pairs} our results for $\epsilon_{b\bf k}-2\epsilon_{{\bf k}_m}$ with the binding energy of two holes found in previous numerical studies as $E_2-2E_1+E_0$, where $E_i$ is the ground-state energy of the system with $i$ holes. It is seen from Fig.~\ref{pairs} that in agreement with our findings both exact diagonalization studies and valence Monte Carlo (VMC) calculations show a threshold value of $J$ above which the binding energy is negative and it is energetically favorable for holes to form couples (Cooper pairs). The value of this threshold obtained in our consideration is in a quantitative agreement with VMC results whereas an extrapolation of exact diagonalization results to the thermodynamic limit seems to give slightly smaller threshold. It should be noted also that all numerical data suffers from finite size effects that may be one of the reasons why they differ from each other.

Superconducting order parameters of different symmetries are built as linear combinations of operators 
\begin{equation}
\label{delta}
	\Delta_{ij} = c_{{\bf r}_i\uparrow} c_{{\bf r}_j\downarrow},
\end{equation}
where $i$ and $j$ are neighboring sites. \cite{leerev,ogatarev,grosrev,scalapair} It is seen from Eqs.~\eqref{trans2} that $\Delta_{ij}$ does not have linear terms in its representation if $i$ and $j$ do not belong to the same unit cell. In contrast, it can be shown that the representation of $\Delta_{ij}$ built as it is described above does contain terms linear in $b^\dagger$ operators when $i$ and $j$ belong to the same unit cell:
\begin{equation}
	\Delta_{ij}  \sim
	\frac{\sqrt n}{2} \cos\alpha 
	\left( (\cos\alpha -\sin\alpha)b^\dagger_{1i} + b^\dagger_{2i} + b^\dagger_{3i} + (\cos\alpha + \sin\alpha)b^\dagger_{4i}\right).
\end{equation}
Then, one has in the leading order in $1/n$ in the superconducting state $\langle\Delta_{ij}\rangle\sim\sqrt{nN}\rho$ and $\langle\Delta_{ij}\rangle\sim \rho\sqrt{N/n}$ if $i$ and $j$ belong to the same and to different unit cells, respectively, where $\rho$ is the density of the condensate of $b$ quasiparticles. As a result, $1/n$ series for $\langle\Delta_{ij}\rangle$ differ for vertical and horizontal bonds of the lattice that is a consequence of the discrimination of these bonds by the two-site variant of the BOT presented here. Then, the two-site BOT is not convenient for the consideration of the symmetry of the superconducting order parameter related with the condensate of $b$ quasiparticles because it requires going far beyond the first order in $1/n$ to describe it accurately. It will be shown in subsequent works that this inconvenience of the theory does not arise in a four-site variant of the BOT built on the extended unit cell having the form of a plaquette.

\section{Conclusion and outlook}
\label{disc}

To conclude, we present the bond-operator theory (BOT) for analytical consideration of the $t$--$J$ model and its extensions with longer-range hopping terms. The BOT is based on the representation of electron operators via localized spins $1/2$ and spinless Fermi-operators of mobile holons. This representation which was originally proposed in Ref.~\cite{sb1} and refined in Ref.~\cite{sb2} allows to accurately get rid of constraint \eqref{cons} of the $t$--$J$ model. The main ingredient of the BOT is the subsequent representation of operators of spins, holons, and electrons containing a rich zoo of Bose- and Fermi-operators which act in the Hilbert space of all quantum states of the whole unit cell. We provide a general scheme of construction of such representation for arbitrary number of sites in (possibly extended) unit cell. The proposed technique is distinguished from previously suggested approaches by a combination of the following features. 

(i) The proposed BOT provides a regular expansion of physical quantities in powers of $1/n$ using conventional diagrammatic technique, where $n\ge1$ is the maximum number of introduced quasiparticles (both bosons and fermions) which can occupy a unit cell. 

(ii) The suggested representation reproduces the commutation algebra of all operators appearing in the theory at any $n>0$. 

(iii) BOT allows to consider both magnetically ordered and disordered (with a singlet ground state) phases and transitions between them. 

(iv) "Physical" and "unphysical" subspaces containing, respectively, no more than $n$ and more than $n$ quasiparticles in the unit cell are not mixed in the BOT. 

(v) Although the expansion parameter $1/n$ is not small in the physically meaningful case of $n=1$, a good {\it quantitative} agreement with previous numerical findings arises at $n=1$ even in the first order in $1/n$ after taking into account a few simple diagrams shown in Fig.~\ref{diag}. 

(vi) Some elementary excitations described in the BOT by separate bosons or fermions appear in common approaches as bound states of conventional quasiparticles. In particular, there are two-fermion bound states (Cooper pairs of two holes) which are described within the BOT by separate bosons. Then, it is possible to find their spectra in the first orders in $1/n$ by taking into account a few simple diagrams, rather than by summing up some infinite series of diagrams as in conventional approaches.

To demonstrate the capabilities of the BOT, we discuss in detail in the present paper the $t$--$J$ model with no more than two holes (magnetic polarons) in the particular case of the square lattice with two lattice sites in the (magnetic) unit cell. Corresponding Fermi- and Bose-operators are introduced in Table~\ref{statestab} and representations of operators appearing in the model are given by Eqs.~\eqref{trans} and \eqref{trans2}. In this case, there are three bosons in the theory describing purely magnetic excitations two of which correspond to spin-1 magnons and the third describes the spin-0 amplitude (Higgs) mode (which appears, e.g, in conventional spin-wave theory as a bound state of two magnons). The magnon spectrum shown in Fig.~\ref{magnons}, the ground-state energy (Eq.~\eqref{eval}), and the staggered magnetization (Eq.~\eqref{mval}) which are obtained in the first order in $1/n$ agree well at $n=1$ with previous numerical, experimental and analytical findings.

As it is seen from  Figs.~\ref{spec03}, \ref{spec04}, and \ref{spectttp}, spectrum of the lower energy fermionic branch and its spectral weight in the electron Green's function obtained in the first order in $1/n$ for different model parameters agree quantitatively at $n=1$ with previous numerical results for magnetic polaron. Besides, our self-consistent calculations of diagrams of the first order in $1/n$ makes this agreement better. The self-consistent procedure converges after 3--5 steps depending on the $J$ value. Interestingly, we obtain dark fermion modes whose spectral weights in the electron Green's function is exactly zero. However they lie substantially above the polaron branch.

We show that the superconducting order parameter built as a linear combination of operators of the form \eqref{delta} is expressed via bosonic operators $b$ from the two-holon sector. All branches of $b$ bosons are overdamped due to their decay into two fermions described by diagram shown in Fig.~\ref{diag}(l) except for the lower branch which becomes well defined around $\bf k=0$ at $J\agt0.8$. We show that $\epsilon_{b\bf k}-2\epsilon_{{\bf k}_m}$, where $\epsilon_{b\bf k}$ is the spectrum of the lower boson branch and $\epsilon_{{\bf k}_m}$ is the polaron energy at its minimum, corresponds at $\bf k=0$ to the binding energy $E_b$ of two holes examined before numerically and it agrees well with previous numerical findings of $E_b$ at $J\agt0.8$ (see Fig.~\ref{pairs}).

Despite some success obtained in the present consideration in describing low-energy properties of the $t$--$J$ model, the two-site variant of the BOT used here is inconvenient, e.g., for consideration of the symmetry of the superconducting order parameter. This is related with the fact that BOT discriminates lattice bonds inside unit cells and between them (see Fig.~\ref{system}). This inconvenience will be overcome in subsequent studies devoted to the four-site variant of the BOT built on the extended unit cell having the form of a plaquette. Besides, it will be shown that unlike the two-site BOT considered here, dark fermion modes appear in the four-site BOT which lie near the polaron branch. 

Notice that the above described general procedure of the construction of operators representations can be applied also for the simplest case of one site in the unit cell. We develop the one-site formalism in Appendix~\ref{method1} and show that the obtained representation is very similar to that derived in Ref.~\cite{chang} from other considerations and that self-consistent calculations are required in the one-site technique to achieve sufficiently good quantitative agreement with previous numerical results for polaron properties.

\section*{Data availability}

Data will be provided by the author upon reasonable request.

\begin{acknowledgments}

This work is supported by the Russian Science Foundation (Grant No.\ 26-22-00364). 

\end{acknowledgments}

\appendix

\section{Additional representations of operators}
\label{oprep}

Using the general procedure described in Sec.~\ref{method2}, we construct the following representations of operators appearing in Eqs.~\eqref{cop} and \eqref{s} in order to substitute them directly to Hamiltonian \eqref{ham0}:
\begin{subequations}
\label{trans2}
\begin{eqnarray}
\label{s1+2}
h_1 h_1^\dagger S_{1}^+ &=& \left(h_1 h_1^\dagger S_{1}^-\right)^\dagger = \cos\alpha P a_{2} - \sin\alpha a_{1}^\dagger P 
+ \cos\alpha \left( a_{1}^\dagger a_{3} + e_{1}^\dagger e_{3} + e_{2}^\dagger e_{4} \right)\nonumber\\
&&{} 
+ \sin\alpha \left( a_{3}^\dagger a_{2} + e_{4}^\dagger e_{3} - e_{2}^\dagger e_{1} \right),\\
\label{s1z2}
h_1 h_1^\dagger S_{1}^z &=& n\frac{\cos2\alpha}{2} 
+\frac{\sin 2\alpha}{2} \left(P a_{3} + a_{3}^\dagger P\right) 
+ \frac12\left( a_{1}^\dagger a_{1} - a_{2}^\dagger a_{2} 
+ e_{2}^\dagger e_{2} - e_{3}^\dagger e_{3} \right)
+\frac{\sin 2\alpha}{2} \left(e_{1}^\dagger e_{4} + e_{4}^\dagger e_{1}\right)
\nonumber\\
&&{}
- \frac{\cos2\alpha}{2}\left( a_{1}^\dagger a_{1} + a_{2}^\dagger a_{2} + 2a_{3}^\dagger a_{3} 
+ e_{2}^\dagger e_{2} + e_{3}^\dagger e_{3} + 2e_{4}^\dagger e_{4} +\sum_{i=1}^4\sum_{r=b,d} r_i^\dagger r_i \right),\\
h_2 h_2^\dagger S_{2}^+ &=& \left(h_2 h_2^\dagger S_{2}^-\right)^\dagger = 
\cos\alpha a_{1}^\dagger P - \sin\alpha P a_{2} 
+ \cos\alpha \left( a_{3}^\dagger a_{2} + d_{2}^\dagger d_{1} + d_{4}^\dagger d_{3} \right)\nonumber\\
&&{} + \sin\alpha \left( a_{1}^\dagger a_{3} + d_{2}^\dagger d_{4} - d_{1}^\dagger d_{3} \right),\\
\label{s2z2}
h_2 h_2^\dagger S_{2}^z &=& -n\frac{\cos2\alpha}{2} 
-\frac{\sin 2\alpha}{2} \left(P a_{3} + a_{3}^\dagger P\right) 
+ \frac12\left( a_{1}^\dagger a_{1} - a_{2}^\dagger a_{2} 
+ d_{2}^\dagger d_{2} - d_{3}^\dagger d_{3} \right)
-\frac{\sin 2\alpha}{2} \left(d_{1}^\dagger d_{4} + d_{4}^\dagger d_{1}\right)
\nonumber\\
&&{}
+ \frac{\cos2\alpha}{2}\left( a_{1}^\dagger a_{1} + a_{2}^\dagger a_{2} + 2a_{3}^\dagger a_{3} 
+ d_{2}^\dagger d_{2} + d_{3}^\dagger d_{3} + 2d_{4}^\dagger d_{4} +\sum_{i=1}^4\sum_{r=b,e} r_i^\dagger r_i\right),\\
\label{ss2}
h_1 h_1^\dagger \left({\bf S}_{1}{\bf S}_{2}\right)h_2 h_2^\dagger &=& - n^2\frac{1+2\sin2\alpha}{4} 
+ n\frac{\cos 2\alpha}{2} \left(P a_{3} + a_{3}^\dagger P\right) 
+ n\sin2\alpha a_{3}^\dagger a_{3} \nonumber\\
&&{}+ n\frac{1+\sin2\alpha}{2}\left(a_{1}^\dagger a_{1} + a_{2}^\dagger a_{2} \right)
+ n\frac{1+2\sin2\alpha}{4} \sum_{i=1}^4 \sum_{r=b,d,e} r_i^\dagger r_i,\\
\label{cu2}
{\mathfrak c}_{1\uparrow} &=& \frac{1}{\sqrt {2n}} \left(
\cos\alpha\left( d_1^\dagger \cos\alpha + d_3^\dagger + d_4^\dagger \sin\alpha \right)P 
+ d_2^\dagger a_1 + b_1^\dagger e_1 + b_2^\dagger e_2
+ \sin^2\alpha \left( d_4^\dagger a_3- b_1^\dagger e_1 + b_4^\dagger e_4 \right)
\right.
\nonumber\\
&&{} + \frac{\sin2\alpha}{2} \left( d_1^\dagger a_3 + b_4^\dagger e_1 + b_1^\dagger e_4 \right)
+ \cos\alpha \left( d_4^\dagger a_1 + b_3^\dagger e_1 + b_4^\dagger e_2 \right)
\nonumber\\
&&{}\left.
+ \sin\alpha \left( d_3^\dagger a_3 - d_1^\dagger a_1 + b_3^\dagger e_4 - b_1^\dagger e_2 \right)
\right),\\
{\mathfrak c}_{2\uparrow} &=& \frac{1}{\sqrt {2n}} \left(
\sin\alpha\left( e_1^\dagger \sin\alpha - e_3^\dagger - e_4^\dagger \cos\alpha \right)P 
+ e_2^\dagger a_1 - b_1^\dagger d_1 - b_2^\dagger d_2
+ \cos^2\alpha \left( e_4^\dagger a_3 + b_1^\dagger d_1 - b_4^\dagger d_4 \right)
\right.
\nonumber\\
&&{} + \frac{\sin2\alpha}{2} \left( b_4^\dagger d_1 + b_1^\dagger d_4 - e_1^\dagger a_3 \right)
+ \cos\alpha \left( e_1^\dagger a_1 + e_3^\dagger a_3 - b_1^\dagger d_2 - b_3^\dagger d_4 \right)
\nonumber\\
&&{}\left.
+ \sin\alpha \left( e_4^\dagger a_1 + b_3^\dagger d_1 - b_4^\dagger d_2 \right)
\right),\\
{\mathfrak c}_{1\downarrow} &=& \frac{1}{\sqrt {2n}} \left(
\sin\alpha\left( d_1^\dagger \sin\alpha - d_2^\dagger - d_4^\dagger \cos\alpha \right)P 
+ d_3^\dagger a_2 + b_1^\dagger e_1 + b_3^\dagger e_3
+ \cos^2\alpha \left( d_4^\dagger a_3- b_1^\dagger e_1 + b_4^\dagger e_4 \right)
\right.
\nonumber\\
&&{} - \frac{\sin2\alpha}{2} \left( d_1^\dagger a_3 + b_4^\dagger e_1 + b_1^\dagger e_4 \right)
+ \cos\alpha \left( d_1^\dagger a_2 + d_2^\dagger a_3 + b_1^\dagger e_3 + b_2^\dagger e_4 \right)
\nonumber\\
&&{}\left.
+ \sin\alpha \left( d_4^\dagger a_2 - b_2^\dagger e_1 + b_4^\dagger e_3 \right)
\right),\\
\label{cd2}
{\mathfrak c}_{2\downarrow} &=& \frac{1}{\sqrt {2n}} \left(
\cos\alpha\left( e_1^\dagger \cos\alpha + e_2^\dagger + e_4^\dagger \sin\alpha \right)P 
+ e_3^\dagger a_2 - b_1^\dagger d_1 - b_3^\dagger d_3
+ \sin^2\alpha \left( e_4^\dagger a_3 + b_1^\dagger d_1 - b_4^\dagger d_4 \right)
\right.
\nonumber\\
&&{} + \frac{\sin2\alpha}{2} \left( e_1^\dagger a_3 - b_4^\dagger d_1 - b_1^\dagger d_4 \right)
+ \sin\alpha \left( e_2^\dagger a_3 - e_1^\dagger a_2 + b_1^\dagger d_3 - b_2^\dagger d_4 \right)
\nonumber\\
&&{}\left.
+ \cos\alpha \left( e_4^\dagger a_2 - b_2^\dagger d_1 - b_4^\dagger d_3 \right)
\right),\\
{\mathfrak c}_{1\uparrow}^\dagger {\mathfrak c}_{2\uparrow}^{} &=& \frac{1}{2} \left(
\sin^2\alpha e_4^\dagger d_1 - \cos^2\alpha e_1^\dagger d_4 - e_2^\dagger d_2
- \cos\alpha \left( e_1^\dagger d_2 + e_2^\dagger d_4 \right)
+ \frac{\sin2\alpha}{2} \left( e_1^\dagger d_1 - e_4^\dagger d_4 \right)
\right.
\nonumber\\
&&{} \left.
+ \sin\alpha \left( e_2^\dagger d_1 - e_4^\dagger d_2 \right) 
\right),\\
{\mathfrak c}_{1\downarrow}^\dagger {\mathfrak c}_{2\downarrow}^{} &=& \frac{1}{2} \left(
\sin^2\alpha e_1^\dagger d_4 - \cos^2\alpha e_4^\dagger d_1 - e_3^\dagger d_3
- \cos\alpha \left( e_3^\dagger d_1 + e_4^\dagger d_3 \right)
+ \frac{\sin2\alpha}{2} \left( e_1^\dagger d_1 - e_4^\dagger d_4 \right)
\right.
\nonumber\\
&&{} \left.
+ \sin\alpha \left( e_1^\dagger d_3 - e_3^\dagger d_4 \right) 
\right),
\end{eqnarray}
\end{subequations}
where $P$ is given by Eq.~\eqref{proj}.

\section{Technique for one site in the unit cell}
\label{method1}

We develop in this section a one-site formalism based on Fermi-Bose representations of operators ${\bf S}_j$, $h_j$, and ${\mathfrak c}_{j\sigma}$ introduced in Sec.~\ref{constr}. Notice that we adopt the general scheme for construction of these representations which we use in Sec.~\ref{method2}.

\subsection{Representation of operators}

Let us define a vacuum state at each lattice site $j$
\begin{equation}
\label{0}
|j0\rangle = |j\!\downarrow \rangle_s |j0\rangle_h
\end{equation}
and one Bose-operator $a_j$ and two Fermi-operators $e_{1j}$ and $e_{2j}$ which create the remaining three states from ${\cal H}_j^s\otimes{\cal H}_j^h$ as
\begin{eqnarray}
\label{opdef}
	a_j^\dagger|j0\rangle &=& |a_j\rangle = \left|j\!\uparrow \rangle_s \right. |j0\rangle_h,\nonumber\\
	e_{1j}^\dagger|j0\rangle &=& |e_{1j}\rangle = 
	\frac{1}{\sqrt2} \left(\left|j\!\downarrow \rangle_s + \left|j\!\uparrow \rangle_s \right) \right.\right. |j1\rangle_h,\\
	e_{2j}^\dagger|j0\rangle &=& |e_{2j}\rangle = 
	\frac{1}{\sqrt2} \left(\left|j\!\downarrow \rangle_s - \left|j\!\uparrow \rangle_s \right) \right.\right. |j1\rangle_h.\nonumber
\end{eqnarray}
Representation of operators ${\bf S}_{j}$, $h_{j}$, and ${\mathfrak c}_{j\sigma}$ via operators $a_j$, $e_{1j}$, and $e_{2j}$ can be built according to their action on states $|j0\rangle$, $|a_j\rangle$, $|e_{1j}\rangle$, and $|e_{2j}\rangle$, the result being 
\begin{eqnarray}
\label{rep10}
	S^z_j &=& -\frac n2 + a_j^\dagger a^{}_j
	+ \frac12 \left(e_{1j}^\dagger - e^\dagger_{2j}\right) \left(e^{}_{1j} - e^{}_{2j}\right),\nonumber\\
	S_j^+ &=& a_j^\dagger P_j + \frac12 \left(e_{1j}^\dagger - e^\dagger_{2j}\right)\left(e^{}_{1j} + e^{}_{2j}\right),\nonumber\\
	S_j^- &=& P_ja^{}_j + \frac12 \left(e_{1j}^\dagger + e^\dagger_{2j}\right)\left(e^{}_{1j} - e^{}_{2j}\right),\\
	h_j   &=& \frac{1}{\sqrt{2n}} P_j \left(e^{}_{1j} + e^{}_{2j}\right) 
	+ \frac{1}{\sqrt{2n}} a_j^\dagger \left(e^{}_{1j} - e^{}_{2j}\right),\nonumber
\end{eqnarray}
where 
\begin{equation}
\label{proj1}
P_j = \sqrt{n - a_j^\dagger a^{}_j - e_{1j}^\dagger e^{}_{1j} - e_{2j}^\dagger e^{}_{2j}}
\end{equation}
and $n=1$. It can be checked that operators in left-hand sides of Eqs.~\eqref{rep10} act on states $|j0\rangle$, $|a_j\rangle$, $|e_{1j}\rangle$, and $|e_{2j}\rangle$ as operators in right-hand sides. Besides, it can be verified straightforwardly that representation \eqref{rep10} reproduces the commutation algebra of all operators for any $n>0$ (i.e., $[S^\sigma_{j},S^\beta_{j}]=i\epsilon_{\sigma\beta\gamma}S^\gamma_{j}$, $[h_{j},{\bf S}_{j}]=0$, $\{h_{j}^\dagger,h_{j}\}=1$, and $\{h_{j},h_{j}\}=0$). Representation \eqref{rep10} shares many properties with representation \eqref{trans} which are discussed in Sec.~\ref{method2}.

As it is done in Sec.~\ref{method2}, it is more convenient and straightforward to derive the following Bose-Fermi representations of operators ${\mathfrak c}_{j\sigma}$ and $h_j h_j^\dagger {\bf S}_j$ given by Eqs.~\eqref{cop} and \eqref{s} and $h_j^\dagger h^{}_j$ in order to substitute them directly to Hamiltonian \eqref{ham0}:
\begin{eqnarray}
\label{rep11}
	h_jh_j^\dagger S_j^z &=& -\frac n2 + a_j^\dagger a_j,\nonumber\\
	h_jh_j^\dagger S_j^+ &=& a_j^\dagger P_j,\nonumber\\
	h_jh_j^\dagger S_j^- &=& P_ja_j,\\
	h_j^\dagger h^{}_j &=& e_{1j}^\dagger e^{}_{1j} + e_{2j}^\dagger e^{}_{2j}, \nonumber\\
	{\mathfrak c}_{j\uparrow} &=& \frac{1}{\sqrt n} e_{1j}^\dagger a_j,\nonumber\\
	{\mathfrak c}_{j\downarrow} &=& \frac{1}{\sqrt n} e_{1j}^\dagger P_j.\nonumber
\end{eqnarray}
It is not surprising that Eqs.~\eqref{rep10} and \eqref{rep11} contain the Holstein-Primakoff spin representation for $S=1/2$ at $n=1$ and $e_{1j}=e_{2j}=0$. Then, boson $a$ describes conventional magnons. Notice also that representation \eqref{rep11} is very similar to that suggested in Ref.~\cite{chang} which was obtained however from other considerations.

\subsection{Hamiltonian transformation}

Notice that representation \eqref{rep11} is valid for one magnetic sublattice if the ground state has the N\'eel order (see Eq.~\eqref{0} for the vacuum state). In order to use Eqs.~\eqref{rep11} for the analysis of the $t$--$J$ model in this case, one has to rotate spin operators by $\pi$ around $x$ axis on sites belonging to the second sublattice. It is easy to show using Eqs.~\eqref{cop} and \eqref{u12} that operators transforms as follows upon this rotation: $S^{y,z}\mapsto-S^{y,z}$, $S^x\mapsto S^x$, ${\mathfrak c}_\uparrow\mapsto{\mathfrak c}_\downarrow$, and ${\mathfrak c}_\downarrow\mapsto{\mathfrak c}_\uparrow$, so that the system Hamiltonian \eqref{ham0} acquires the form
\begin{equation}
\label{ham11}
	{\cal H} 
	= 
	-t\sum_{\langle p,j\rangle \sigma} 
	\left(
	{\mathfrak c}^\dagger_{p\sigma} {\mathfrak c}^{}_{j-\sigma}
	+ {\mathfrak c}^\dagger_{j\sigma} {\mathfrak c}^{}_{p-\sigma} \right)
	+
	J\sum_{\langle p,j\rangle} \left(h^{}_ph^\dagger_p
	\left( S^x_pS^x_j - S^y_pS^y_j - S^z_pS^z_j \right)
	h^{}_jh^\dagger_j
	- \frac14h^\dagger_ph^{}_ph^\dagger_jh^{}_j\right)
	+J\sum_j h^\dagger_jh^{}_j.
\end{equation}

It is shown below that representation \eqref{rep11} allows to find all observable quantities as series in powers of $1/n$ using conventional diagrammatic technique. For this purpose, it is also convenient to perform the formal renormalization in the first and the last terms in Eq.~\eqref{ham11}
\begin{eqnarray}
\label{rent}
	t &\mapsto& nt,\nonumber\\
	J\sum_j h^\dagger_jh^{}_j &\mapsto& nJ\sum_j h^\dagger_jh^{}_j
\end{eqnarray}
after which all constant terms in the Hamiltonian become of the order of $(1/n)^{-2}$, terms linear in Bose and Fermi operators are ${\cal O}((1/n)^{-3/2})$, bilinear terms are of $(1/n)^{-1}$ order, etc. Then, $n$ plays the role very much similar to the spin value $S$ in the Holstein-Primakoff transformation with the only difference that only $n=1$ has the physical meaning. 

Substituting Eqs.~\eqref{rep11} to Eq.~\eqref{ham11}, taking into account the formal renormalization \eqref{rent}, and expanding the square root in operator $P$, one obtains Eq.~\eqref{hbf}, where ${\cal H}_1=0$,
\begin{eqnarray}
	{\cal H}_2 &=& 
	nJ\sum_{\bf k} \left(
	\zeta_{\bf 0} a^\dagger_{\bf k}a^{}_{\bf k} 
	+\frac12\zeta_{\bf k} \left(a^{}_{\bf k}a^{}_{-\bf k} + a^\dagger_{\bf k}a^\dagger_{-\bf k} \right)
	+e_{1\bf k}^\dagger e^{}_{1\bf k} + e_{2\bf k}^\dagger e^{}_{2\bf k}
	\right),\\
	{\cal H}_3 &=& 
	\sqrt n \frac{2}{\sqrt N}\sum_{{\bf k}_1+{\bf k}_2+{\bf k}_3=0} 
	e^\dagger_{1-{\bf k}_1} e^{}_{1{\bf k}_2}
	\left(
	\zeta_{{\bf k}_2} a^{}_{{\bf k}_3}
	+
	\zeta_{{\bf k}_1} a^\dagger_{-{\bf k}_3}
	\right),\\
	{\cal H}_4 &=& 
	-J \frac{1}{2N} \sum_{{\bf k}_1+{\bf k}_2+{\bf k}_3+{\bf k}_4=0} 
	\left(
	a^\dagger_{-{\bf k}_1} 
	\left(2 \zeta_{{\bf k}_2+{\bf k}_3} a^\dagger_{-{\bf k}_2}  a^{}_{{\bf k}_3}
	+ \zeta_{{\bf k}_2} a^{}_{{\bf k}_2} a^{}_{{\bf k}_3} 
	+ \zeta_{{\bf k}_2} a^\dagger_{-{\bf k}_2} a^\dagger_{-{\bf k}_3}
	\right)	a^{}_{{\bf k}_4}
	\right.\nonumber\\
	&&{}
	+\zeta_{{\bf k}_1} \left(a^{}_{{\bf k}_1} a^{}_{{\bf k}_2}+a^\dagger_{-{\bf k}_1} a^\dagger_{-{\bf k}_2}\right) 
	\left(e^\dagger_{1-{\bf k}_3} e^{}_{1{\bf k}_4}
	+e^\dagger_{2-{\bf k}_3} e^{}_{2{\bf k}_4}\right)\nonumber\\
	&&\left.{}
	+\frac12 
		\left(e^\dagger_{1-{\bf k}_1} e^{}_{1{\bf k}_2}
	+e^\dagger_{2-{\bf k}_1} e^{}_{2{\bf k}_2}\right)
		\left(e^\dagger_{1-{\bf k}_3} e^{}_{1{\bf k}_4}
	+e^\dagger_{2-{\bf k}_3} e^{}_{2{\bf k}_4}\right)
\right),
\end{eqnarray}
$N$ is the number of sites in the lattice now,
\begin{equation}
	\zeta_{\bf k} = \cos k_1+\cos k_2
\end{equation}
and $k_p=({\bf k}\mbox{\boldmath $\mathfrak w$}_p)$ are components of $\bf k$ ($p=1,2$, see Fig.~\ref{system}(b)). 

\subsection{Diagrammatic consideration}

\begin{figure}
\includegraphics[scale=0.4]{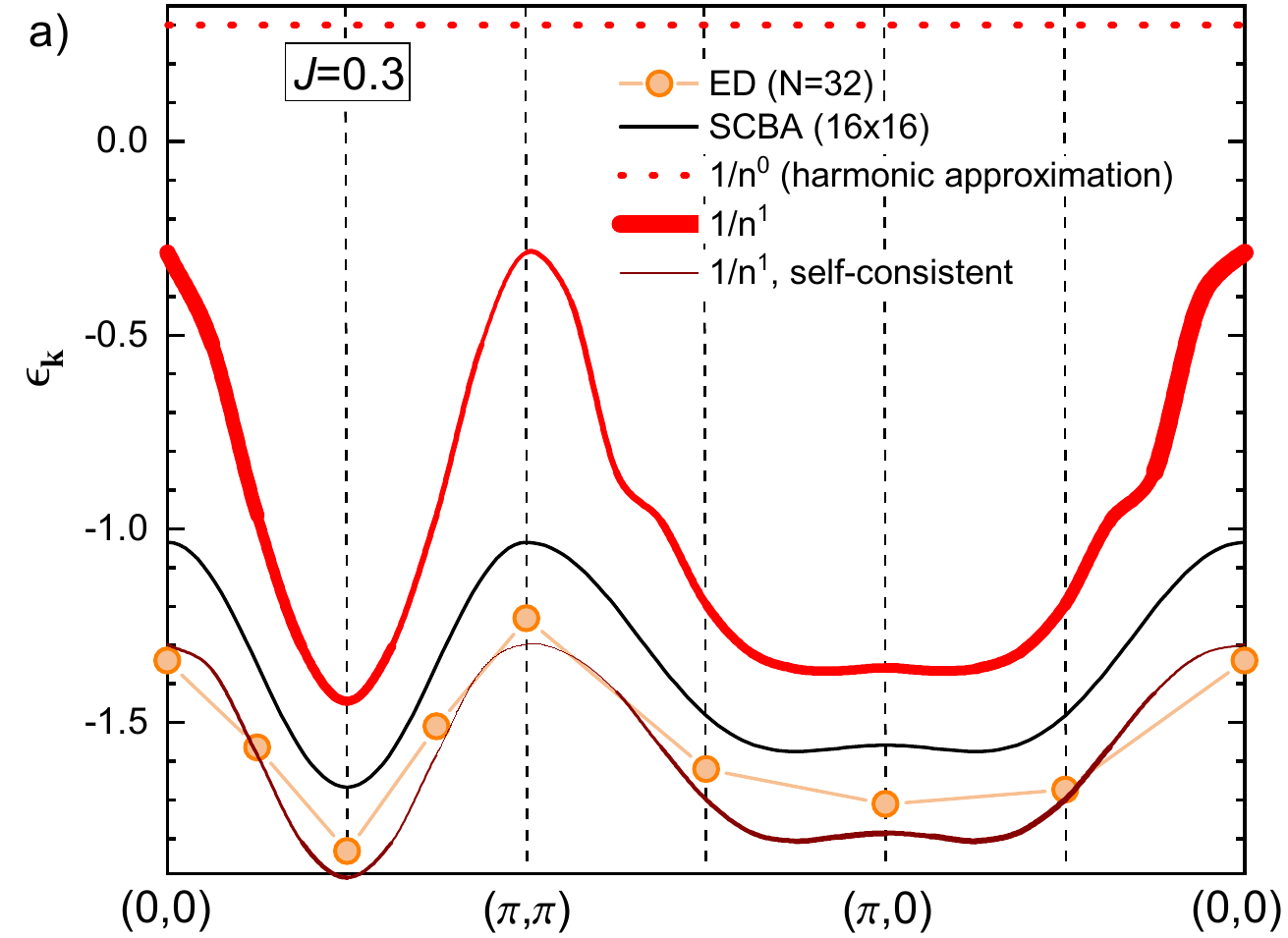}
\includegraphics[scale=0.4]{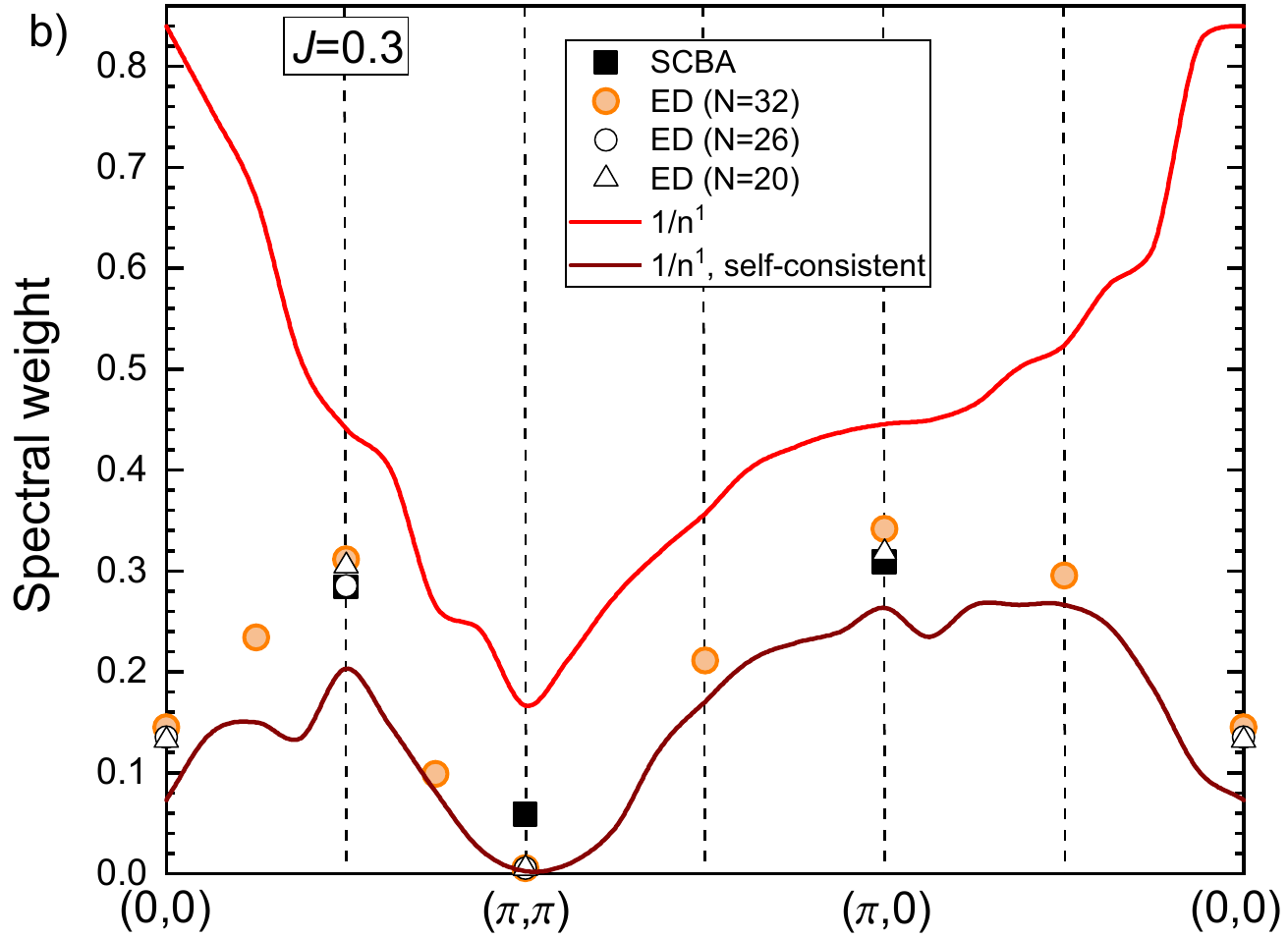}
\caption{
Same as Fig.~\ref{spec03} but with our results obtained within the one-site formalism.
\label{spec031site}}
\end{figure}

\begin{figure}
\includegraphics[scale=0.4]{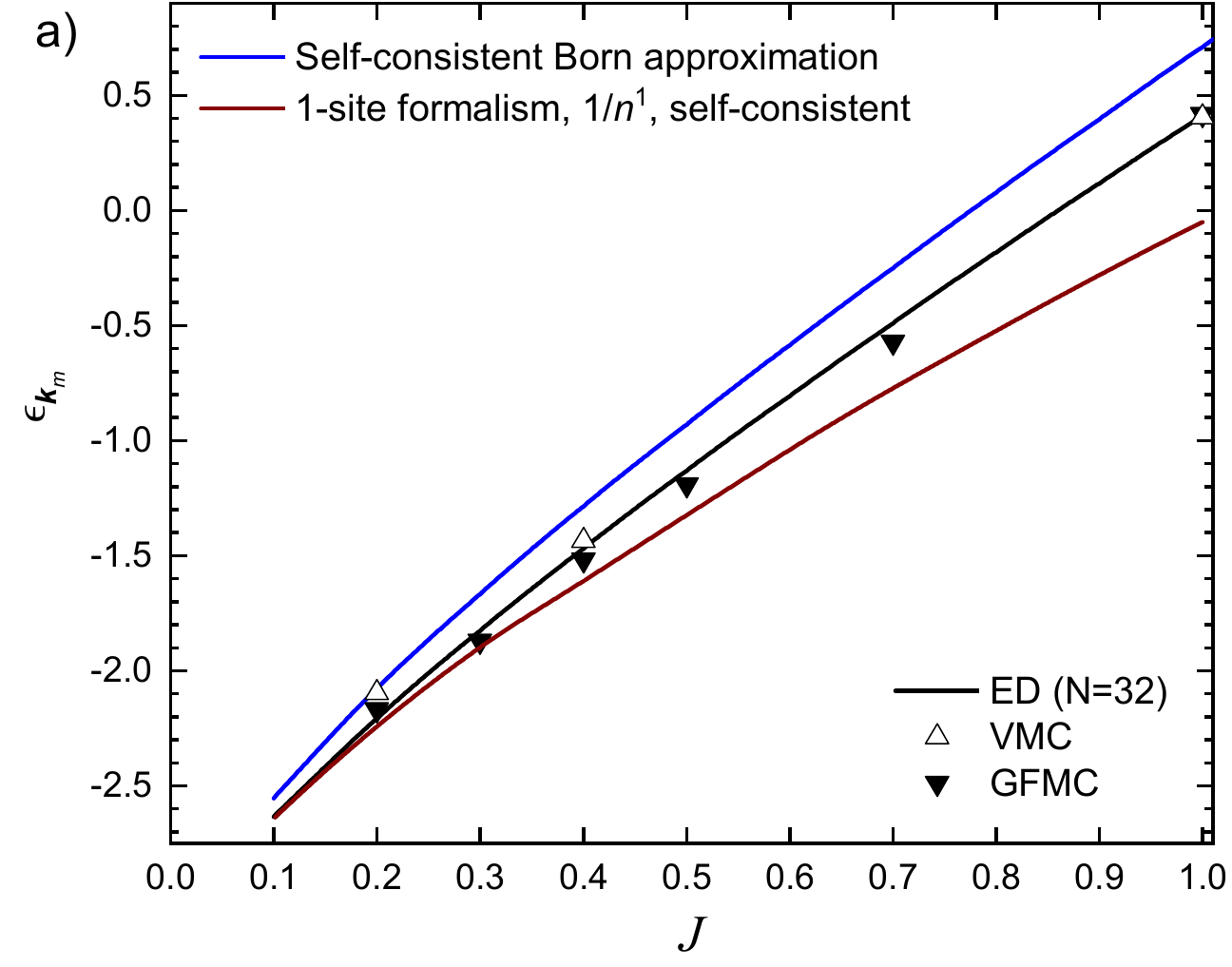}
\includegraphics[scale=0.4]{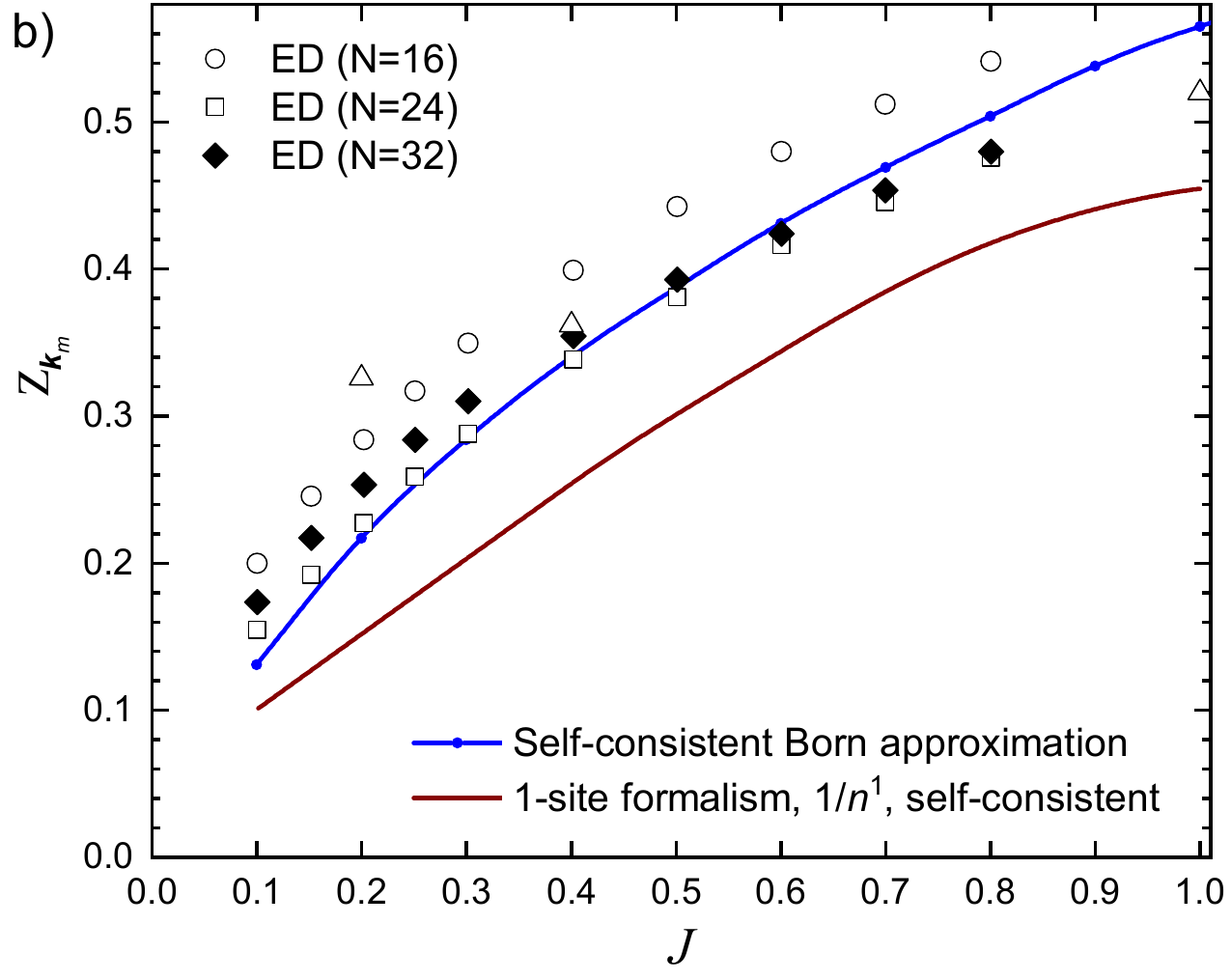}
\includegraphics[scale=0.4]{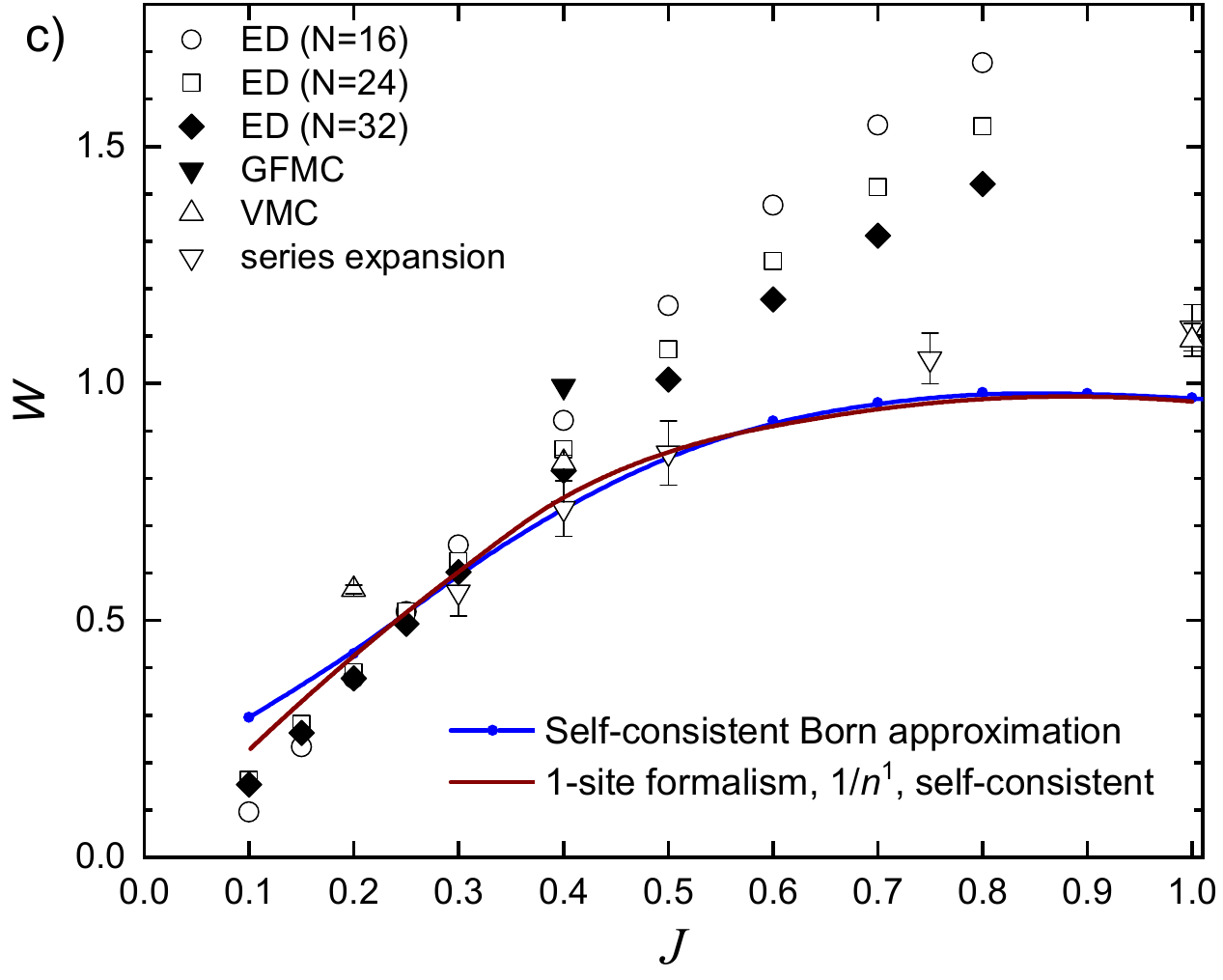}
\caption{
Our findings obtained self-consistently within the one-site formalism are shown together with the same numerical data as in Fig.~\ref{ez} for the polaron energy $\epsilon_{{\bf k}_m}$, the polaron spectral weight $Z_{{\bf k=k}_m}$, and the polaron bandwidth $W$.
\label{ez1site}}
\end{figure}

Diagrammatic consideration can be carried out as it is done in Sec.~\ref{hafbi}. In particular, one has diagrams shown in Figs.~\ref{diag}(a)--\ref{diag}(i) in the first order in $1/n$. Renormalization of magnons at zero concentration of holes within the one-site formalism is completely equivalent to that in the spin-wave theory based on the Holstein-Primakoff transformation. Our results for the polaron spectrum at $J=0.3$ are presented in Fig.~\ref{spec031site}. It is seen that, in contrast to the results obtained in the two-site formalism (Fig.~\ref{spec03}), the first order in $1/n$ is definitely insufficient for a quantitative description of polaron properties although self-consistent calculations within the first order in $1/n$ carried out as it is explained in Sec.~\ref{method2} give a good agreement with numerical findings. Fig.~\ref{ez1site} shows that self-consistent calculations give quantitatively correct results for polaron properties at $J=0.1\div1$.

\bibliography{htsbib}

\end{document}